\documentclass[a4paper,fleqn]{cas-sc}

\usepackage[authoryear,longnamesfirst]{natbib}
\usepackage{graphicx}
\usepackage{subcaption}
\usepackage{caption}

\def\tsc#1{\csdef{#1}{\textsc{\lowercase{#1}}\xspace}}
\tsc{WGM}
\tsc{QE}
\tsc{EP}
\tsc{PMS}
\tsc{BEC}
\tsc{DE}
\begin{document}
\let\WriteBookmarks\relax
\def\floatpagepagefraction{1}
\def\textpagefraction{.001}

\shorttitle{Leveraging social media news}

\shortauthors{CV Radhakrishnan et~al.}

\title [mode = title]{Can LLM Improve for Expert Forecast Combination? Evidence from the European Central Bank Survey}                      
\tnotemark[1,2]

% \tnotetext[1]{This document is the results of the research
%    project funded by the National Science Foundation.}

% \tnotetext[2]{The second title footnote which is a longer text matter
%    to fill through the whole text width and overflow into
%    another line in the footnotes area of the first page.}

% \author[1,3]{CV Radhakrishnan}[type=editor,
                        % auid=000,bioid=1,
                        % prefix=Sir,
                        % role=Researcher,
                        % orcid=0000-0001-7511-2910]

% \cormark[1]


% \ead{cvr_1@tug.org.in}

% \ead[url]{www.cvr.cc, cvr@sayahna.org}

\author[label1,label2]{Yinuo Ren} %% Author name

\author[label1,label2]{Jue Wang\corref{cor1}}
\cortext[cor1]{Corresponding author}
\affiliation[label1]{organization={MOE Social Science Laboratory of Digital Economic Forecasts and Policy Simulation, University of Chinese Academy of Sciences},%Department and Organization
            city={Beijing},
            postcode={100190}, 
            country={China}}
            
\affiliation[label2]{organization={AMSS Center for Forecasting Science, Chinese Academy of Scienc},%Department and Organization
            city={Beijing},
            postcode={100190}, 
            country={China}}

\begin{abstract}
This template helps you to create a properly formatted \LaTeX\ manuscript.

\noindent\texttt{\textbackslash begin{abstract}} \dots 
\texttt{\textbackslash end{abstract}} and
\verb+\begin{keyword}+ \verb+...+ \verb+\end{keyword}+ 
which
contain the abstract and keywords respectively. 

\noindent Each keyword shall be separated by a \verb+\sep+ command.
\end{abstract}

% \includegraphics{figs/grabs.pdf}

% \begin{highlights}
% \item Research highlights item 1
% \item Research highlights item 2
% \item Research highlights item 3
% \end{highlights}

% Keywords

\maketitle

\section{Introduction}

Professional forecasters play a critical role in generating macroeconomic predictions that guide economic decision-making. However, the inherent uncertainty involved in forecasting complex economic variables underscores the need for adopting forecast combination methods. The European Central Bank (ECB) introduced its Survey of Professional Forecasters (SPF) in 1999 to systematically collect macroeconomic predictions for the euro area. This initiative focuses on experts from financial and non-financial institutions within the European Union, gathering forecasts for key economic indicators such as the Harmonized Index of Consumer Prices (HICP) inflation, real Gross Domestic Product (GDP) growth, and unemployment rates across various time horizons \citep{Garcia2003introduction}.
SPF contributes significantly to monetary policy decisions by offering insights into private sector inflation expectations, which are essential for understanding economic dynamics. While it does not replace the Eurosystem's own analysis, the survey provides a complementary perspective that strengthens the ECB's assessment of risks to price stability \citep{Carlos2007}.

Even professional forecasters often struggle with accuracy due to external macroeconomic uncertainty and internal overconfidence, emphasizing the need for forecast combinations to address these challenges \citep{CASEY2021716} \citep{LI2021103266}. The theoretical foundation of forecast combination can be traced back to Bates and Granger \citep{Bates01121969}, who derived optimal weights by minimizing the Mean Squared Error (MSE) of two base models. Despite significant advances in combination methods, the "Forecast Combination Puzzle" reveals that simple averaging often outperforms more complex machine learning algorithms \citep{Stock}. Empirical evidence supports the consistent effectiveness of the equal-weighted strategy across different time periods, target variables, and forecast horizons \citep{GENRE2013108}. 

However, existing forecast combination methods face several critical challenges. First, the estimation of the covariance matrix is crucial for determining the optimal weights, but this becomes unreliable with a small sample size in the SPF situation. The time-varying nature of covariance matrix risks overfitting and lag phenomenon \citep{burgi2017nonparametric}. Secondly, missing observations in unbalanced panels can introduce additional uncertainty when imputation methods are used \citep{lahiri2017online}. The third problem is about interpretability. Optimal weights often lack clear explanations and are unable to effectively capture nonlinear patterns \citep{kruger2017survey}.

Recent advancements in Generative Artificial Intelligence (AI) and Large Language Models (LLMs) are revolutionizing various domains, including time series forecasting \citep{jin2023time,liu2024timecma}, financial analysis \citep{chen2024does}, and medical research \citep{yuan2024continued,garcia2024medical}. State-of-the-art LLMs such as Generative Pre-trained Transformer(GPT) \citep{roumeliotis2023chatgpt}, Llama \citep{touvron2023llama}, and Deepseek \citep{bi2024deepseek} demonstrate several key advantages that make them particularly suitable for addressing the challenges in SPF data. At their core, LLMs possess strong generalization capability that enable zero-shot learning without extensive training data. Building on this foundation, they are able to effectively handle missing values with comprehensive prior knowledge. Furthermore, unlike traditional machine learning methods that often operate as "black boxes", LLMs provide detailed explanations of decision-making processes \citep{makridakis2023large}. 

These inherent advantages position LLMs as particularly suitable for ensemble learning. However, most existing research merely employs LLMs as base learners, combining outputs through voting schemes or other traditional methods. For example, in medical question answering, multiple LLMs are used collaboratively to improve response accuracy and reliability \citep{Sivara} \citep{maharjan2024openmedlm} \citep{singhal2025toward}. In information security, ensembles of LLMs work together to identify and fix code bugs \citep{pan2023risk}. While LLMs show promise in forecast combinations, their performance, especially in the context of SPF, remains largely undetermined. Therefore, comparing the ensemble capabilities of LLM against simple averaging emerges as a crucial research question. 

This study aims to assess the performance of LLMs in forecast combination using a zero-shot learning approach, meaning the models do not require fine-tuning or additional network information \citep{kojima2022large}. By leveraging the LLM's ability to recognize patterns, we input fixed-window historical data on expert performance, allowing the models to detect and understand each expert's prediction patterns and biases. The LLM then automatically determines the best combination of weights based on its understanding of expert behavior, without the need for supervised training. We apply this zero-shot combination framework across various scenarios, such as different macroeconomic indicators, periods of high and low disagreement, and varying levels of attentiveness. This comprehensive analysis allows us to compare the effectiveness of LLM-based intelligent ensemble methods with simple averaging in terms of forecast accuracy and robustness.

The findings of our study contribute to both the forecast 
combination and the emerging field of LLM applications 
in economic forecasting. Our study makes the following key contributions:
\begin{itemize}
    \item We propose an approach by directly employing LLMs for forecast combinations in expert surveys, moving beyond the conventional paradigm of treating LLMs merely as base models.
    \item We demonstrate the effectiveness of zero-shot learning, where LLMs can dynamically adjust combination weights without requiring extensive training data or parameter tuning.
    \item We conduct comprehensive scenario analyses examining LLM performance across different macroeconomic indicators, expert disagreement levels, and attention variations, providing insights into the robustness and adaptability of LLM-ensemble.
    % 异质性检验
\end{itemize}
Our research also offers practical insights for the ECB on leveraging advanced AI technologies to enhance its forecasting capabilities. The remainder of this paper is structured as follows. Section 2 develops our research hypotheses. Section 3 specifies our empirical model. Section 4 presents the empirical analysis and results. Section 5 concludes.

\section{Hypothesis development}

The hypotheses in this study aim to evaluate the performance of LLMs relative to simple averaging (SA) in forecast combination across various conditions. Actual values are included as control variables to account for the potential confounding effects of macroeconomic volatility, as market shocks may lead experts to adopt more conservative forecasts. By isolating the effects of key variables, we ensure that the results more accurately capture the true impacts, rather than being influenced by fluctuations in actual values.

\subsection{Comparative Accuracy of LLM-Ensemble and Simple Averaging (H1)}

Simple equally weighted pooling of forecasts has demonstrated strong empirical performance in the Survey of Professional Forecasters \citep{GENRE2013108}. However, this approach exhibits inherent limitations, including vulnerability to market shocks and lag issues. Expert predictions frequently exhibit high correlations, manifesting as simultaneous underestimation or overestimation patterns \citep{radchenko2023too}. In contrast, LLMs utilize pre-trained pattern recognition algorithms to analyze historical windows of individual experts, enabling dynamic weight assignment \citep{silver2016mastering}. 

The comparative efficacy between simple averaging and LLM-ensemble presents a compelling empirical question. While experts can demonstrate rapid adaptation to new information, their effectiveness is often constrained by psychological biases and limited capacity for processing a large amount of information. Leveraging LLMs' ability to discern trends and patterns from historical data, we hypothesize that they will provide more accurate forecast combinations than simple averaging.

\textbf{Hypothesis 1 (H1):} \textit{LLMs will surpass simple averaging in forecast combination, when adjusted for comparable levels of actual values.}

\subsection{Comparative Advantage of LLM-Ensemble in Different Indicators (H2)}

\begin{figure}[!htbp]
    \centering
    \begin{subfigure}{0.48\textwidth}
        \includegraphics[width=\textwidth]{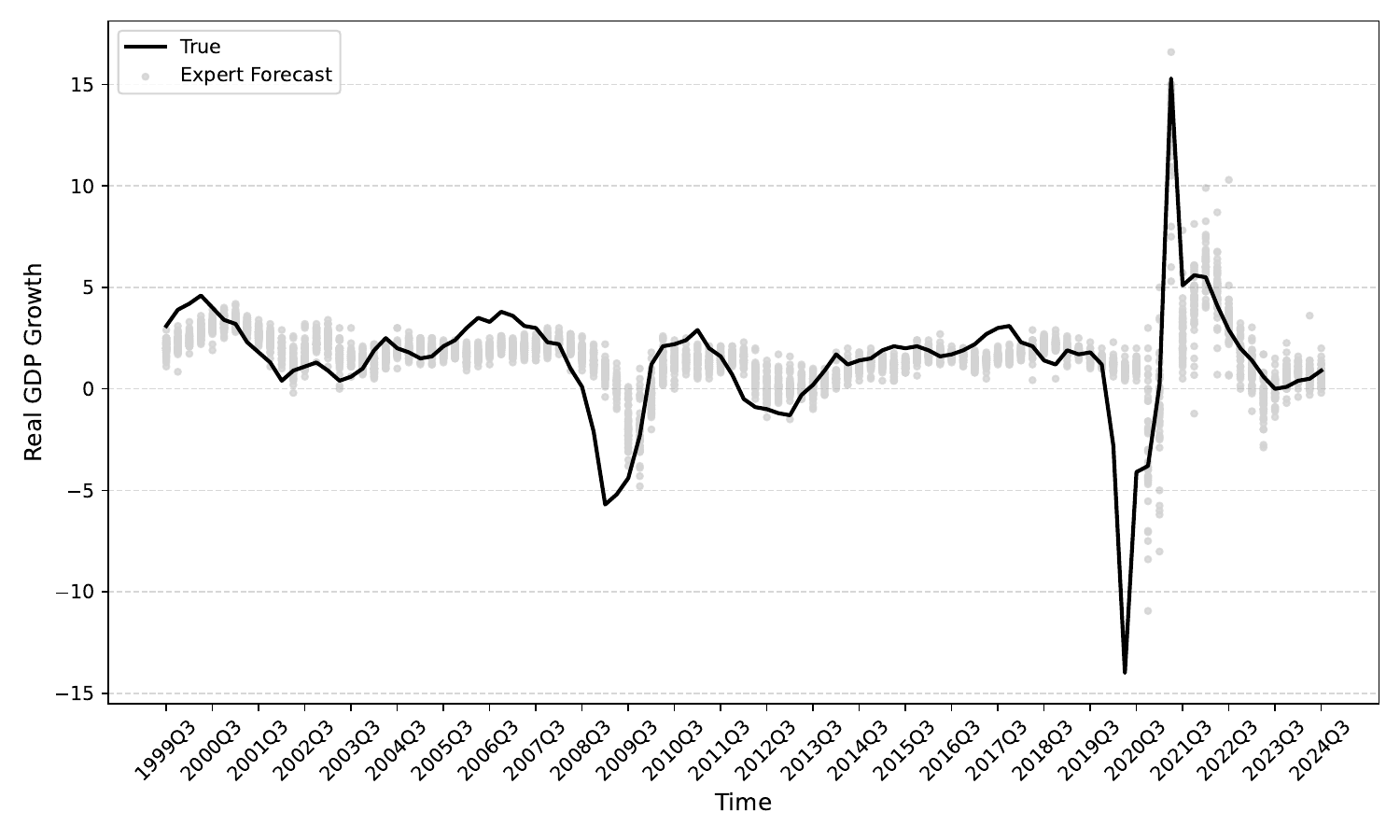}
        \caption{Real GDP Growth (h=1)}
        \label{fig:GDP_1}
    \end{subfigure}
    \hfill
    \begin{subfigure}{0.48\textwidth}
        \includegraphics[width=\textwidth]{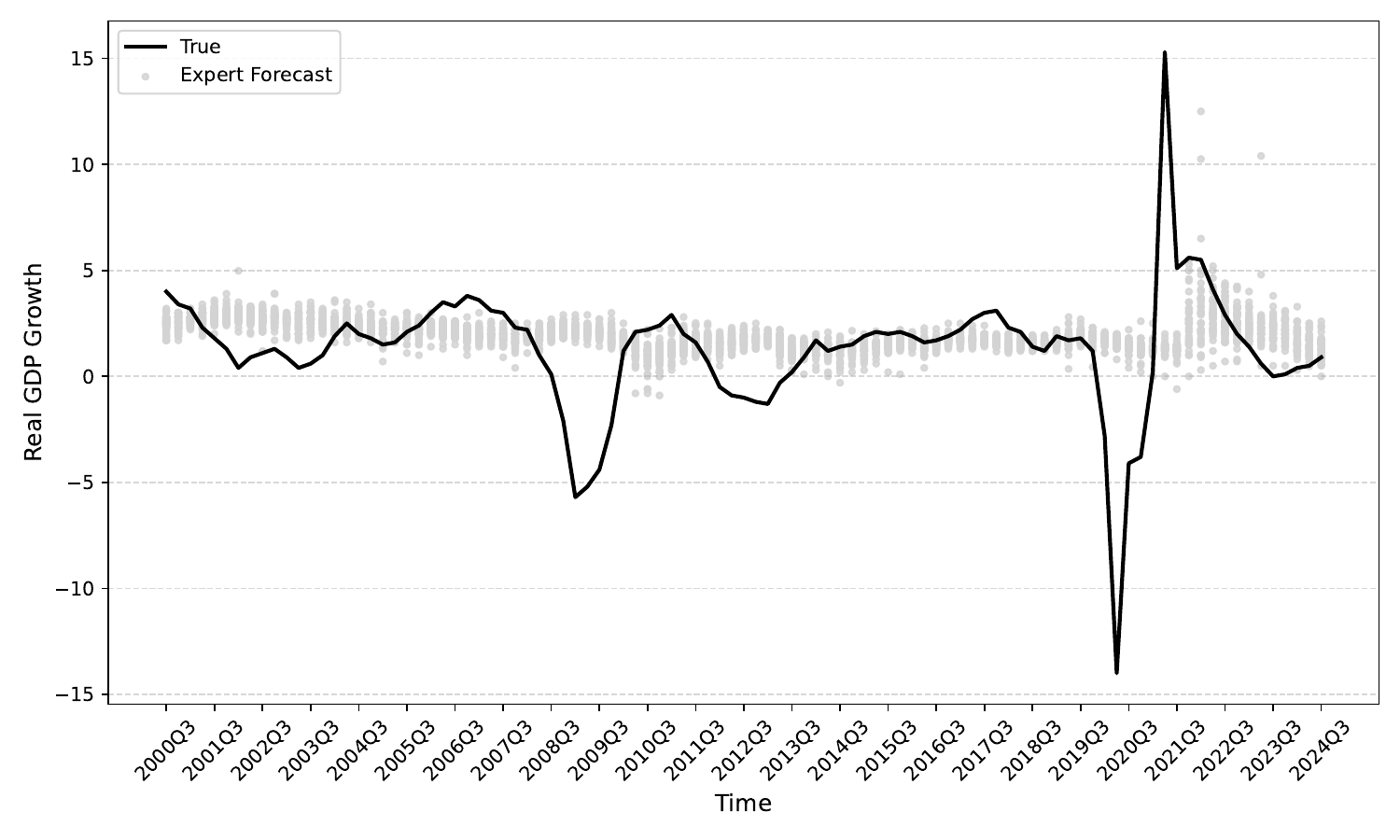}
        \caption{Real GDP Growth (h=2)}
        \label{fig:GDP_2}
    \end{subfigure}
    
    \begin{subfigure}{0.48\textwidth}
        \includegraphics[width=\textwidth]{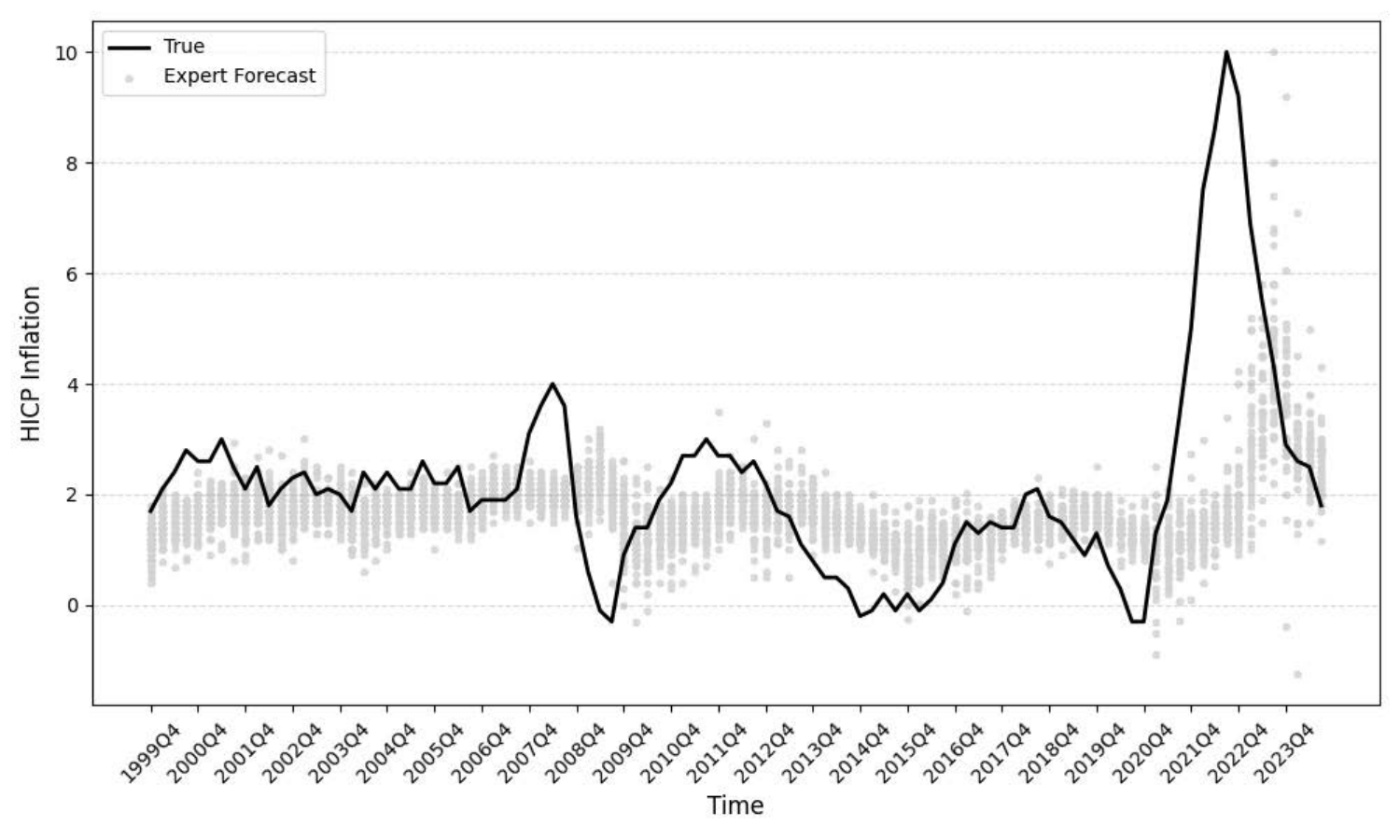}
        \caption{HICP Inflation (h=1)}
        \label{fig:Inflation_1}
    \end{subfigure}
    \hfill
    \begin{subfigure}{0.48\textwidth}
        \includegraphics[width=\textwidth]{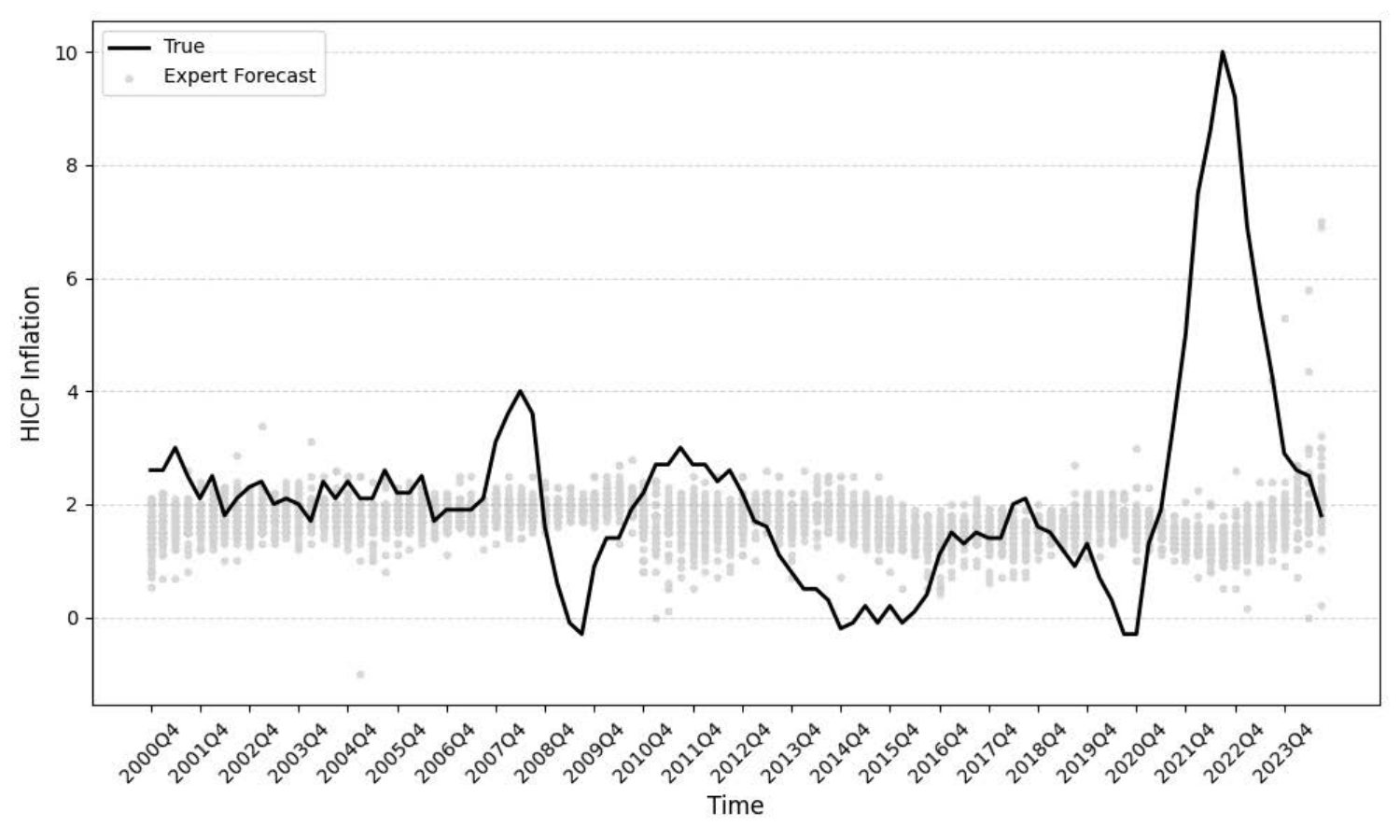}
        \caption{HICP Inflation (h=2)}
        \label{fig:Inflation_2}
    \end{subfigure}
    
    \begin{subfigure}{0.48\textwidth}
        \includegraphics[width=\textwidth]{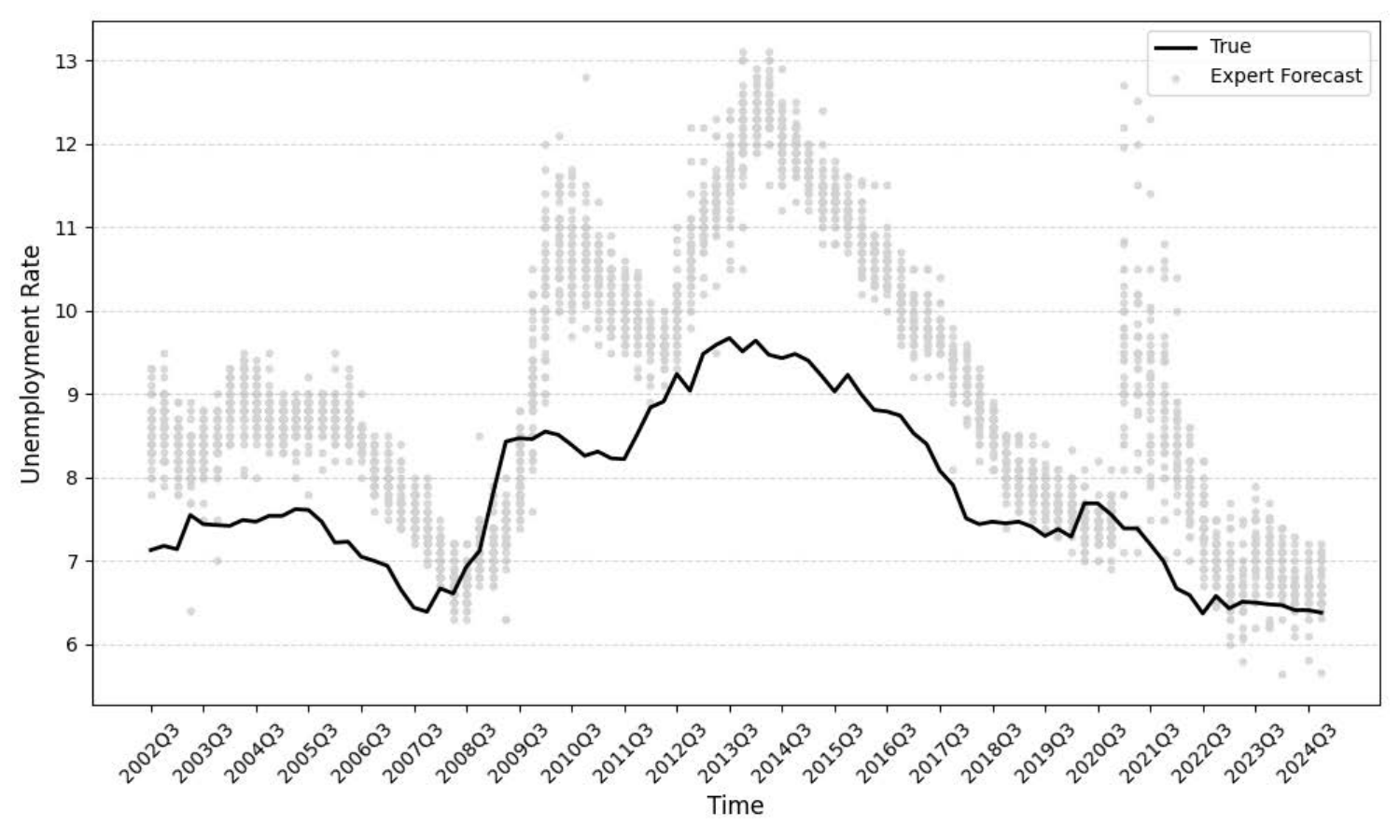}
        \caption{Unemployment Rate (h=1)}
        \label{fig:Unempolyment_1}
    \end{subfigure}
    \hfill
    \begin{subfigure}{0.48\textwidth}
        \includegraphics[width=\textwidth]{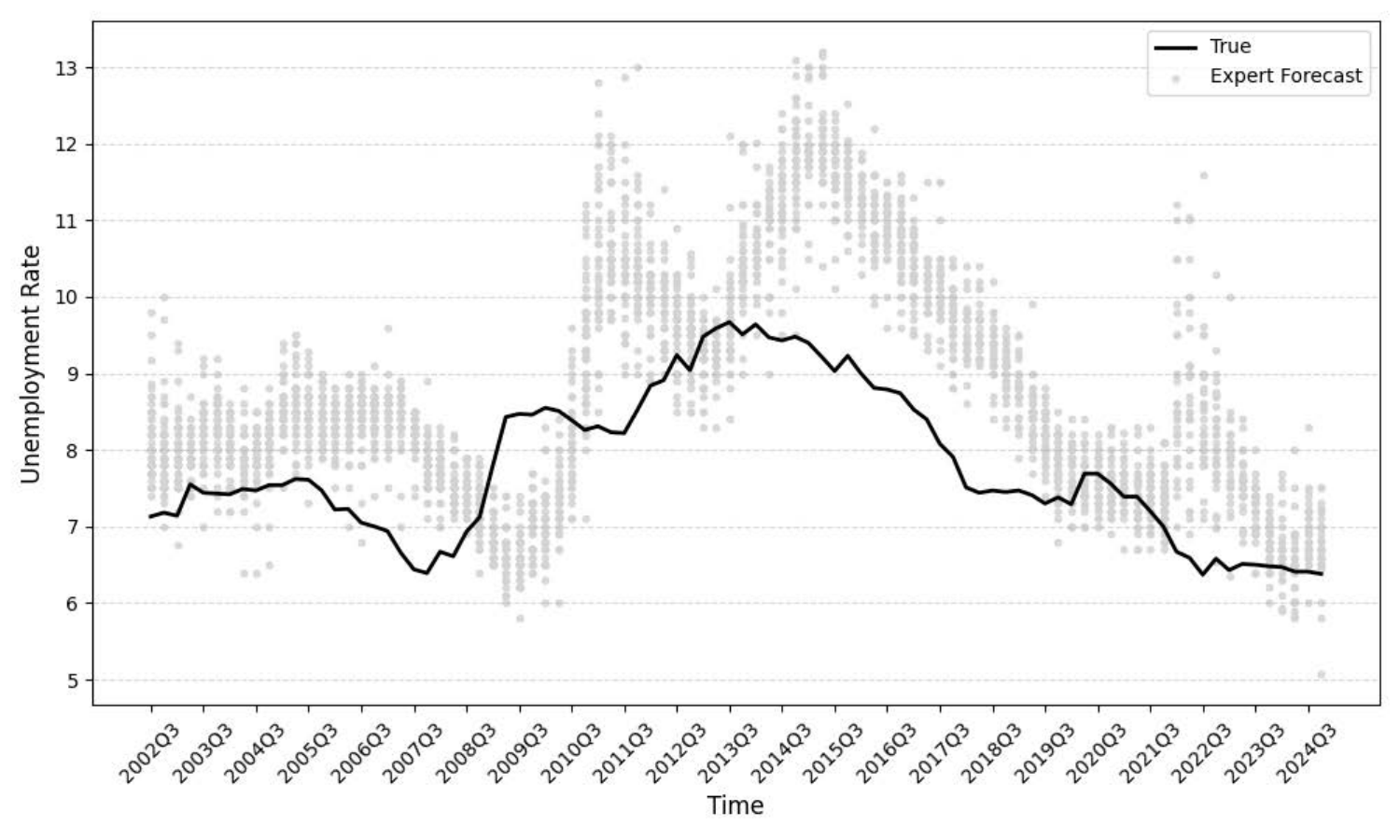}
        \caption{Unemployment Rate (h=2)}
        \label{fig:Unempolyment_2}
    \end{subfigure}
    \caption{One-year and Two-year Forecasts of Three Macroeconomic Indicators.}
    \label{fig:forecasts}
\end{figure}

Figure \ref{fig:forecasts} illustrates the quarterly survey data for three macroeconomic indicators, with solid lines representing actual values and gray scatter points indicating forecast values. A notable observation is the pronounced lag effect in expert predictions, where forecast adjustments typically follow market fluctuations with a delay. Among the three indicators, real GDP Growth forecasts demonstrate the most comprehensive coverage of actual values, followed by HICP inflation, while unemployment rate predictions exhibit the poorest performance. This variation reflects the inherent challenges and differential predictive capabilities across these economic indicators. 

Based on these observations, we propose the following hypotheses:

\textbf{Hypothesis 2 (H2):} \textit{LLMs will outperform simple averaging in forecast combination when expert forecasts exhibit high prediction errors, as LLMs can effectively identify and account for poor forecasting performance patterns.}

\subsection{Impact of Expert Disagreement on Forecast Combinations (H3)}

The level of disagreement among professional forecasters serves as a crucial indicator of forecast quality and market uncertainty. Economic shocks tend to amplify forecast dispersion as experts interpret the impact of unexpected events differently, resulting in increased prediction errors \citep{czudaj2022heterogeneity}. However, this relationship may not be linear, as moderate disagreement can sometimes reflect healthy diversity in forecasting approaches. To measure the degree of disagreement $\sigma_t$, we calculate the cross-sectional standard deviation of expert forecasts:

\begin{align*}
\sigma_t = \sqrt{\frac{1}{N-1}\sum_{i=1}^{N}(f_{i,t}-\bar{f}_t)^2},
\end{align*}
where $N$ is the number of experts, $f_{i,t}$ is the forecast value of the $i$-th expert in period $t$, $\bar{f}_t$ is the mean value across all experts in period $t$. Higher values indicate more severe disagreement among experts during this period.

LLMs possess the capability to analyze the distribution of expert predictions and identify patterns in disagreement. They can detect whether high disagreement stems from genuine uncertainty or systematic biases in expert judgment. This nuanced understanding allows LLMs to adjust combination weights more effectively than simple averaging, which treats all forecasts equally regardless of their relative reliability.

\textbf{Hypothesis 3 (H3):} \textit{Both forecasting approaches will experience reduced accuracy under conditions of high expert disagreement (H3a).LLM ensemble will demonstrate superior performance in reconciling divergent expert opinions due to its systematic integration capabilities and immunity to cognitive biases (H3a).}

\subsection{Impact of Expert Inattentiveness on Forecast Combinations (H4)}
We measure forecaster inattentiveness as the proportion of experts who submit identical forecasts across consecutive quarters, indicating a lack of active forecast revision \citep{ANDRADE2013967}. This metric captures forecaster inertia, where higher values suggest more diligent updating based on new information, while lower values may indicate reduced engagement in the forecasting process. 
The degree of inattentiveness is computed as :

\begin{align}
\lambda_{t} = \frac{1}{n_t} \sum_{i=1}^{n_t} I(f_{i,t}(y) = f_{i,t-1}(y)),
\end{align}
where $n_t$ is the number of experts in period $t$, $f_{i,t}(y_{t+h})$ is the forecast value of the $i$-th expert in period $t$ for variable $y$ at horizon $h$, and $I(\cdot)$ is an indicator function equal to 1 if the condition is true and 0 otherwise.
\begin{figure}[htbp]
    \begin{subfigure}{\textwidth}
        \includegraphics[width=\textwidth]{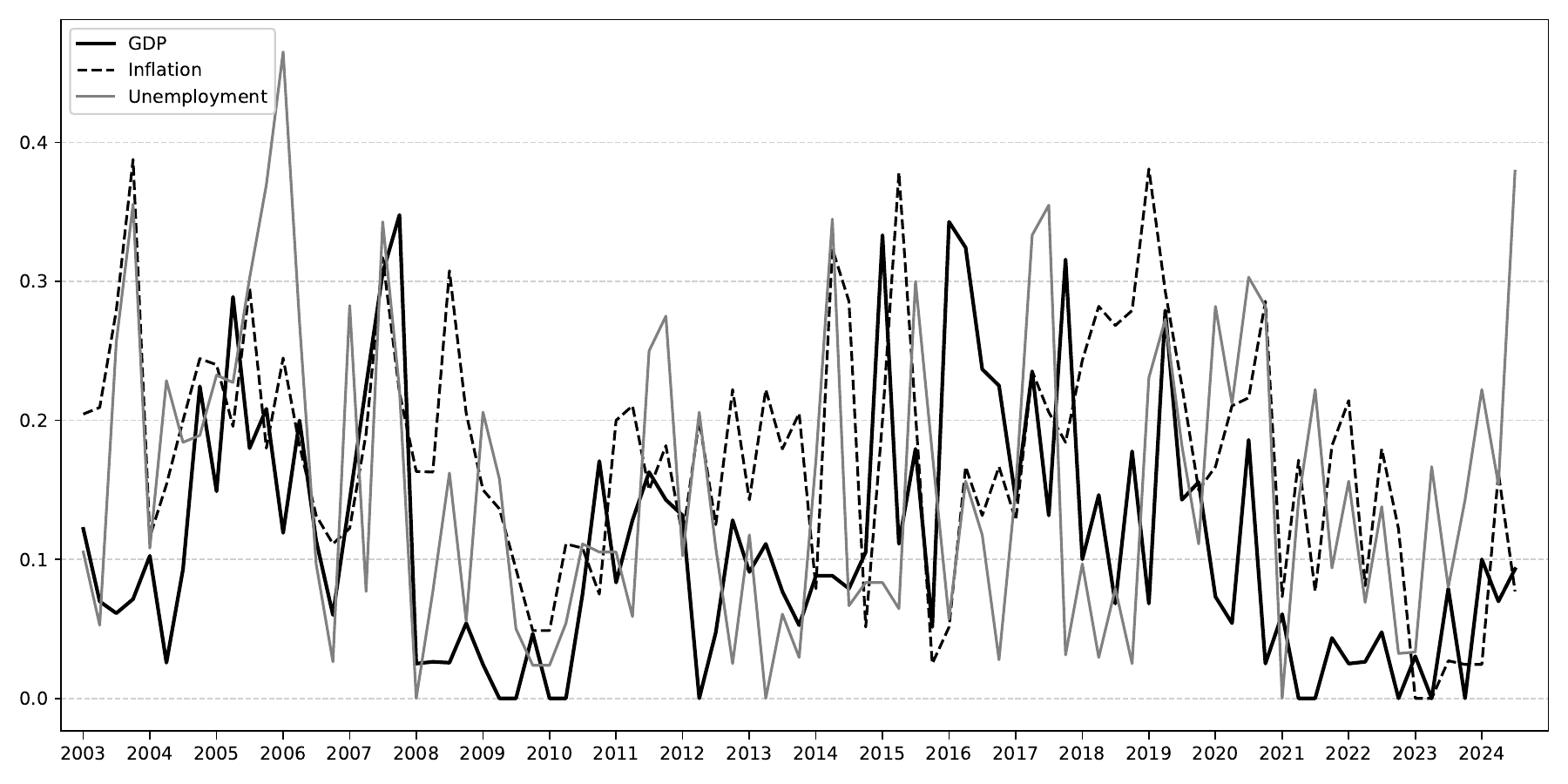}
        \caption{One-year Ahead Forecasts}
        \label{fig:Att_1}
    \end{subfigure}
    
    \begin{subfigure}{\textwidth}
        \includegraphics[width=\textwidth]{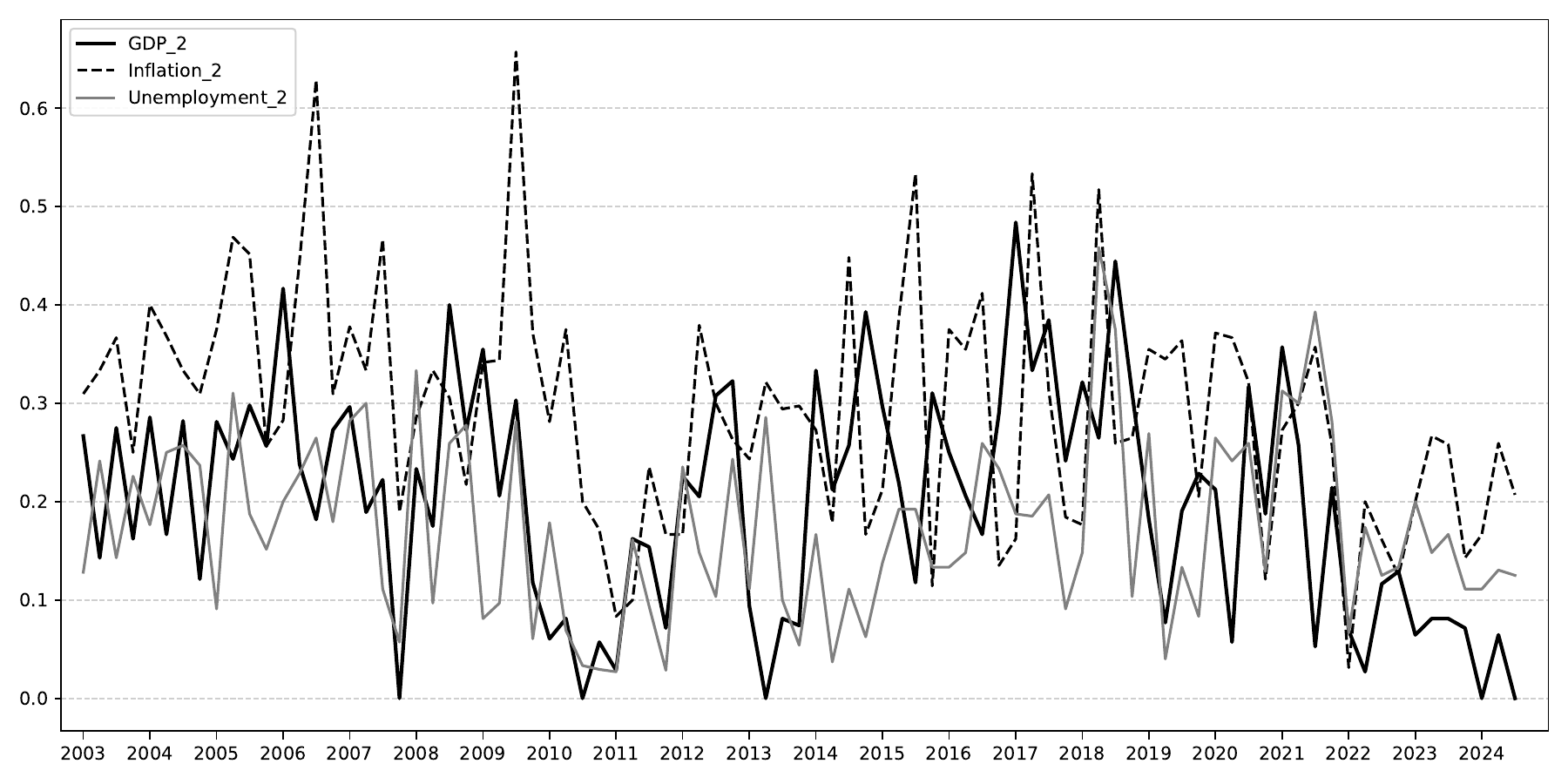}
        \caption{Two-year Ahead Forecasts}
        \label{fig:Att_2}
    \end{subfigure}
    \caption{Evolution of Forecaster Inattentiveness Over Time. }
    \label{fig:inattentives}
\end{figure}

Figure \ref{fig:inattentives} illustrates the evolution of forecaster inattentiveness across different macroeconomic indicators and horizons. For real GDP Growth predictions, experts consistently demonstrate high levels of attentiveness, with low proportions of unchanged forecasts across both one-year and two-year horizons. The unemployment rate exhibits notably different patterns across horizons - there is a high degree of forecast inertia in one-year ahead predictions, while two-year forecasts show improved attentiveness with more frequent revisions.The HICP inflation forecasts display an opposite trend, with experts showing greater attentiveness in short-term forecasts compared to long-term.

In contrast, LLMs can analyze experts' historical performance to identify their forecasting styles and diligence in making predictions. By recognizing which experts consistently provide timely and accurate updates versus those who are less attentive, LLMs can optimally adjust combination weights to enhance the overall forecast accuracy. This systematic approach to weight allocation helps mitigate the impact of varying levels of expert engagement in the forecasting process:

\textbf{Hypothesis 4 (H4):} \textit{The accuracy of expert forecasts will be positively correlated with the level of forecaster attention (H4a), while LLM ensemble performance will demonstrate greater resilience to fluctuations in expert attention levels (H4b).}

\section{Methodology}
Our methodology employs mixed-effects regression models to systematically evaluate the comparative advantages of LLM-ensemble over simple averaging \citep{cnaan1997using}. The model specification allows us to account for both fixed and random effects, capturing the complex dynamics in forecast combinations. Additionally, the effectiveness of our approach heavily relies on prompt engineering - carefully crafted prompts that enable LLMs to generate forecast combinations. A key advantage of our framework is its zero-shot nature, requiring no pre-training or parameter tuning, making it readily deployable for real-world applications.

\subsection{Model Specification}

To test our hypotheses empirically, the general specification of our econometric model is:

\begin{align}
log(Y_{ij})=\beta_0+\beta_1X_{j}+\beta_2Z_{i}+\beta_3U_{ij}+\varepsilon_{j}
\end{align}

Where:

\begin{itemize}
  \item $Y_{ij}$ represents the Absolute Percentage Error (APE) for ensemble method $i$ (where $i=0$ denotes simple averaging and $i=1$ denotes LLM-ensemble) within the $j$-th period. We expect $Y_{ij}$ to be positive, with lower values indicating better forecast accuracy.
  \item $\beta_0$ is the intercept term, expected to be positive and statistically significant at the 1\% level.
  \item $X_{j}$ represents the true value at the $j$-th period. 
  \item $\beta_1$ is hypothesized to be positive and statistically significant at the 5\% level, reflecting the systematic impact of variable magnitude on forecast errors.
  \item $Z_{i}$ is a binary indicator for forecaster type (where $i=0$ denotes simple averaging and $i=1$ denotes LLM-ensemble). 
  \item $\beta_2$ represents the coefficient for the forecaster type indicator, expected to be negative and significant, indicating LLM-ensemble's superior performance.
  \item $U_{ij}$ captures the categorical fixed effects from various experimental conditions, including: indicators (real GDP, HICP inflation, Unemployment), expert disagreement levels (low vs high), inattentiveness impact (negative vs positive) and their respective interactions.
  \item $\beta_3$ represents the coefficients for these fixed effects, with varying signs and significance levels depending on the specific condition being tested.
  \item $\varepsilon_{j}$ denotes the residual error term clustered at the year level, assumed to be normally distributed with mean zero.
\end{itemize}

In our analysis, we calculate the APE as:

\begin{align}
Y_{ij} = \frac{|F_{ij} - X_j|}{X_j},
\end{align}
where $F_{ij}$ represents the forecast value from method $i$ in period $j$, and $X_j$ is the actual realized value. The true value $X_j$ is included as a covariate to control for the magnitude effect of different macroeconomic indicators on prediction errors. The experimental conditions $U_{ij}$, including indicator types (real GDP, HICP inflation, unemployment), expert disagreement levels (low vs high), and inattentiveness impact (negative vs positive), are incorporated as fixed effects. This structure enables us to systematically examine how different contexts influence forecast accuracy. By clustering standard errors at the year level, our model accounts for potential temporal correlation in forecast errors, thus providing more robust statistical inference. This comprehensive approach not only captures the complex dynamics in macroeconomic indicators but also enhances the reliability and generalizability of our empirical findings.

\subsection{Prompt Engineering}
Prompt engineering serves as a fundamental methodology in natural language processing (NLP), enabling effective utilization of LLMs for forecast combination. By carefully crafting specific instructions and queries, we guide Qwen to process and interpret expert forecasts while accounting for individual forecasting patterns and styles \citep{bai2023qwen} \citep{yang2024qwen2}. Our prompt engineering framework consists of three key components:

\begin{itemize}
    \item \textbf{Historical Accuracy Weighting:} The prompt instructs LLMs to dynamically weight expert forecasts based on their historical prediction accuracy and overall performance patterns. This mechanism helps identify and appropriately weight consistently accurate forecasters while reducing the influence of less reliable predictions.
    
    \item \textbf{Lag Compensation:} The framework includes explicit instructions for LLMs to detect and calibrate for temporal lags in expert forecasts. This addresses a common issue in SPF where expert predictions tend to lag behind actual changes in true values, with experts typically adjusting their forecasts only after observing shifts in realized values.
    
    \item \textbf{Trend Enhancement:} LLMs are directed to incorporate recent historical trends of true values to correct for prediction errors. This mechanism helps calibrate forecasts by leveraging observed patterns in actual data.
\end{itemize}

The implementation uses carefully structured prompts that maintain strict output formatting to ensure consistency and prevent contamination across different forecasting periods:

\begin{verbatim}
content = f"""
You are provided with ECB forecast data, including historical true values 
and expert prediction sequences for the past three quarters. Please generate 
an ensemble forecast for {date} following these rules:

1. Historical Accuracy Weighting: Consider each expert's prediction accuracy 
   and historical performance for dynamic weighting.

2. Lag Compensation: Detect and calibrate for temporal lags in expert 
   historical performance.

3. Trend Enhancement: Incorporate recent historical value trends.
   
Please generate an ensemble forecast based on the following data 
(expert sets may vary by quarter):
{formatted_data}

Output must strictly follow this format with no additional fields:
Ensemble Result: <value>
"""
\end{verbatim}

This structured approach enables systematic evaluation of expert forecasts while maintaining forecast diversity and minimizing potential biases. The strict output formatting ensures consistent interpretation across different forecasting periods and prevents information leakage between forecasts.

\subsection{Zero-Shot Learning}
The remarkable zero-shot generalisation capability of LLMs can be explained in theory through statistical learning theory. This concept is rooted in Vapnik–Chervonenkis (VC) theory, which established a fundamental generalisation error bound \citep{vapnik1971uniform}:

\begin{equation}
R(h) \leq \hat{R}_S(h) + \sqrt{\frac{\ln(2/\delta) + h \ln(2en/h)}{2n}},
\end{equation}
where $R(h)$ represents the expected risk, $n$ is the sample size, and $\delta$ is the confidence parameter. This bound demonstrates that with sufficient training data, the generalization error can be effectively controlled. For modern LLMs, the key insight is that when the pretraining data size $m$ vastly exceeds the task complexity, the pretraining error term becomes negligible. Shi formalized this for transformer architectures with $L$ layers and attention width $w$ \citep{shi2023trade} :

\begin{equation}
\epsilon_{\text{zsl}} \leq \frac{\text{poly}(L,w)}{\sqrt{m}} + \inf_{f^* \in \mathcal{F}} \|f^* - f_{\text{target}}\|
\end{equation}
where $\mathcal{F}$ represents the model's function class. This theoretical framework explains how LLMs can perform effective forecast combinations without task-specific fine-tuning: their extensive pre-training ensures minimal generalisation error, even on unseen tasks.

\begin{itemize}
    \item \textbf{Adaptive Panel Composition:} Zero-shot learning allows LLMs to adapt to varying numbers and compositions of expert forecasters across different survey periods without retraining. This flexibility is crucial in SPF where the panel of responding experts often changes between periods.
    
    \item \textbf{Pattern Recognition:} The inherent pattern recognition capabilities of LLMs through in-context learning enable them to identify and incorporate complex relationships during different economic trends. Thus, LLMs can discern the distinctive and evolving forecasting behaviors of individual experts. This allows LLMs to flexibly adjust the weighting of each expert over time, capturing shifts in their predictive accuracy and adapting to changing economic conditions.
    
    \item \textbf{Forecast Independence:} Zero-shot learning preserves the independence of individual expert forecasts by processing them without introducing biases. Since the reasoning and decision-making processes of experts are non-public, ensemble methods that rely solely on historical forecast performance are limited. LLMs' independent analytical capabilities allow for stronger integration of expert forecasts, maintaining diversity and capturing different economic perspectives and scenarios.
\end{itemize}

This flexibility enables the system to account for factors that are often difficult to incorporate in traditional statistical approaches.

\section{Experiments Analysis}
Our analysis utilizes forecast data obtained from the ECB SPF program, which gathers expert forecasts four times annually - once per quarter in January, April, July and October. The data collection takes place during the latter half of the opening month each quarter, allowing forecasters to incorporate current economic data into their assessments. Following each survey round, the ECB first releases preliminary findings via their Monthly Bulletin publication (appearing in February, May, August and November), before making comprehensive results and historical records accessible on their SPF online platform \citep{ECB}. This well-structured and open data publication approach forms a solid basis for conducting our empirical investigation.

\subsection{Comparative Accuracy of LLM-Ensemble and Simple Averaging (H1)}
To evaluate H1, we conducted a comparative analysis between the LLM-ensemble and simple averaging of expert forecasts. For one-year-ahead forecasts, Table \ref{tbl:h1_1} shows that the LLM-ensemble achieves significantly better accuracy with a coefficient of -0.121. For two-year-ahead forecasts in Table \ref{tbl:h1_2}, the LLM-ensemble maintains a negative coefficient of -0.063, though not statistically significant. These results suggest that the LLM-ensemble's advantage is more pronounced in shorter-term forecasts, likely due to its ability to process complex near-term patterns more effectively than simple averaging.

\begin{table}[htbp]
\caption{Comparative Forecast Accuracy Analysis (h=1)}
\label{tbl:h1_1}
\begin{tabular}{@{} lccc @{}}
\toprule
& \multicolumn{3}{c}{APE} \\
\cmidrule{2-4}
& Coef. & SE & P-value \\
\midrule
Intercept & -0.766*** & 0.149 & 0.000 \\
Qwen & -0.121** & 0.047 & 0.016 \\
True & -0.155*** & 0.027 & 0.000 \\
\bottomrule
\end{tabular}
\end{table}

\begin{table}[htbp]
\caption{Comparative Forecast Accuracy Analysis (h=2)}
\label{tbl:h1_2}
\begin{tabular}{@{} lccc @{}}
\toprule
& \multicolumn{3}{c}{APE} \\
\cmidrule{2-4}
& Coef. & SE & P-value \\
\midrule
Intercept & -0.691*** & 0.150 & 0.000 \\
Qwen & -0.063 & 0.077 & 0.418 \\
True & -0.178*** & 0.026 & 0.000 \\
\bottomrule
\end{tabular}
\end{table}

\subsection{Comparative Advantage of LLM-Ensemble in Different Indicators (H2)}

The results in Tables \ref{tbl:h2_gdp}, \ref{tbl:h2_inf}, and \ref{tbl:h2_unemp} reveal several patterns. The LLM-ensemble  consistently shows a stronger performance for one-year-ahead forecasts, with the most notable improvement observed in the Growth GDP.. The coefficients of the growth GDP nad the unemployment rate are negative and statistically significant, indicating better accuracy than the simple average. However, for two-year-ahead forecasts, the advantage of the LLM-ensemble becomes weaker and often loses statistical significance. For the HICP inflation, neither method demonstrates a clear edge, as both coefficients are close to zero and not significant.

These findings suggest a common trend: LLM-ensemble is most effective for short-term forecasts and for indicators that are more sensitive to recent information and its benefit diminishes for longer horizons, primarily because the quality of expert forecasts at this longer horizon is relatively low. As a result, the ability of the LLM-ensemble to improve upon these less reliable expert inputs is limited, leading to a lack of significant performance gains compared to the simple averaging.

\begin{table}[htbp]
    \caption{Comparative Forecast Accuracy Analysis of the Grwth GDP}
    \label{tbl:h2_gdp}
    \begin{tabular}{@{} lcccccccc @{}}
    \toprule
    & \multicolumn{4}{c}{h=1} & \multicolumn{4}{c}{h=2} \\
    \cmidrule(lr){2-5} \cmidrule(lr){6-9}
    & Coef. & SE & t-stat & P-val & Coef. & SE & t-stat & P-val \\
    \midrule
    True & -0.155*** & 0.0410 & -3.80 & 0.001 & -0.132* & 0.047 & -2.76 & 0.011 \\
    Qwen & -0.264** & 0.114 &  -2.30 & 0.03 & 0.106 & 0.125 &  0.85 &  0.402 \\
    \midrule
    Intercept & -0.779*** & 0.195 & -3.98& 0.001 & -0.828*** & 0.199 & -4.16 & 0.000 \\
    \bottomrule
    \end{tabular}
    \end{table}
    
    \begin{table}[htbp]
    \caption{Comparative Forecast Accuracy Analysis of the HICP Inflation}
    \label{tbl:h2_inf}
    \begin{tabular}{@{} lcccccccc @{}}
    \toprule
    & \multicolumn{4}{c}{h=1} & \multicolumn{4}{c}{h=2} \\
    \cmidrule(lr){2-5} \cmidrule(lr){6-9}
    & Coef. & SE & t-stat & P-val & Coef. & SE & t-stat & P-val \\
    \midrule
    True & -0.181 & 0.204 & -0.89 & 0.382 & -0.212 & 0.0214 & -0.99  & 0.331 \\
    Qwen & -0.032 & 0.071 & -0.45 & 0.655 & 0.036 & 0.082 & -0.45 &  0.657 \\
    \midrule
    Intercept & -0.599 & 0.541 & -1.11 & 0.279 & -0.532 & 0.2556 & -0.96 & 0.349 \\
    \bottomrule
    \end{tabular}
    \end{table}
    
    \begin{table}[htbp]
    \caption{Comparative Forecast Accuracy Analysis of the Unemployment Rate}
    \label{tbl:h2_unemp}
    \begin{tabular}{@{} lcccccccc @{}}
    \toprule
    & \multicolumn{4}{c}{h=1} & \multicolumn{4}{c}{h=2} \\
    \cmidrule(lr){2-5} \cmidrule(lr){6-9}
    & Coef. & SE & t-stat & P-val & Coef. & SE & t-stat & P-val \\
    \midrule
    True & -0.339** & 0.115 & 2.94 & 0.008 & 0.176 & 0.0127 & 1.39 & 0.179 \\
    Qwen & -0.061** & 0.023 & -2.64 & 0.015 & -0.272 & 0.186 & -1.46 &  0.158 \\
    \midrule
    Intercept & -4.7172 & 0.979 & -4.82 & 0.000 & -3.448*** &  1.045 & -3.30 & 0.003 \\
    \bottomrule
    \end{tabular}
    \end{table}

\subsection{Impact of Expert Disagreement on Forecast Combinations (H3)}

To test H3, we examined how expert disagreement affects the performance of different forecast combination methods. We first calculated expert disagreement as the standard deviation of individual forecasts for each prediction period, and created a binary variable Disagreement that equals 1 when the standard deviation exceeds the median level of disagreement, and 0 otherwise. Tables \ref{tbl:h3_1} and \ref{tbl:h3_2} present the regression results analyzing how forecast accuracy varies with expert disagreement. For one-quarter-ahead forecasts, the significantly negative coefficient on the interaction term Qwen*Disagreement indicates that the LLM-ensemble method maintains its effectiveness in reducing forecast errors even when experts show high levels of disagreement. For two-quarter-ahead forecasts, while the interaction term remains negative, it is not statistically significant. 

\begin{table}[htbp]
\caption{Impact of Expert Disagreement on Forecast Methods (h=1)}
\label{tbl:h3_1}
\begin{tabular}{@{} lcccc @{}}
\toprule
& Coef. & SE & t-stat & P-val\\
\midrule
True & -0.147*** & 0.028 & -5.22 & 0.000\\
Qwen & -0.038 & 0.062 & -0.61 & 0.544\\
Disagreement & 0.403* & 0.205 & 1.96 & 0.061 \\
Qwen*Disagreement & -0.184** & 0.086 & -2.13 & 0.043 \\
\midrule
Intercept & -0.978*** & 0.223 & -4.39 & 0.000 \\
\bottomrule
\end{tabular}
\end{table}

\begin{table}[htbp]
\caption{Impact of Expert Disagreement on Forecast Methods (h=2)}
\label{tbl:h3_2}
\begin{tabular}{@{} lcccc @{}}
\toprule
& Coef. & SE & t-stat & P-val\\
\midrule
True & -0.182*** & 0.028 & -6.61 & 0.000 \\
Qwen & -0.023 & 0.107 & -0.21 & 0.834 \\
Disagreement & 0.147 & 0.152 & 0.97 & 0.341 \\
Qwen*Disagreement & -0.075 & 0.148 & -0.50 & 0.619 \\
\midrule
Intercept & -0.754*** & 0.171 & -4.42 & 0.000 \\
\bottomrule
\end{tabular}
\end{table}

\subsection{Impact of Expert Inattentiveness on Forecast Combinations (H4)}

In the previous section, we define inattentiveness as the case when an expert gives the same forecast as in the previous period. Since expert forecasts are made on a quarterly basis, the time interval is short and macroeconomic variables may not change much. As a result, some experts may adjust their forecasts less often and simply repeat their previous values. This behavior can be seen as a form of "laziness" in forecasting. 

We create a binary variable Inattentiveness, which equals 1 if $lambda_{t}$ is greater than 0.1 (indicating high inattentiveness), and 0 if the difference is less than 0.1 (indicating the forecast has changed substantially from the previous period). Tables \ref{tbl:h4_1} and \ref{tbl:h4_2} show the regression results.

The empirical results indicate that for one-quarter-ahead forecasts, the interaction term of Qwen and inattentiveness does not attain statistical significance. This suggests that the LLM-ensemble fails to demonstrate a discernible advantage in forecast accuracy. Conversely, for two-quarter-ahead forecasts, the interaction term is negative and approaches significance, implying that the LLM-ensemble may outperform sample averaging under longer forecast horizons. A plausible explanation for this phenomenon lies in the nature of quarterly data, where experts might perceive limited necessity to update forecasts frequently, particularly during periods of stable macroeconomic conditions. When a substantial number of experts reiterate prior forecasts, the informational content embedded in the combined forecast diminishes. In such scenarios, the LLM-ensemble framework can leverage historical data and contextual information to extract latent signals, thereby enhancing predictive performance.

\begin{table}[htbp]
\caption{Impact of Expert Inattentiveness on Forecast Combinations(h=1)}
\label{tbl:h4_1}
\begin{tabular}{@{} lcccc @{}}
\toprule
& Coef. & SE & t-stat & P-val \\
\midrule
True & -0.155*** & 0.027 & -5.84 & 0.000 \\
Qwen & -0.160* & 0.080 & -2.00 & 0.056 \\
Inattentiveness & -0.562* & 0.326 & -1.73 & 0.097 \\
Qwen*Inattentiveness & 0.041 & 0.088 & 0.46 & 0.648 \\
\midrule
Intercept & -0.240 & 0.307 & -0.78 & 0.441 \\
\bottomrule
\end{tabular}
\end{table}

\begin{table}[htbp]
\caption{Impact of Expert Inattentiveness on Forecast Combinations (h=2)}
\label{tbl:h4_2}
\begin{tabular}{@{} lcccc @{}}
\toprule
& Coef. & SE & t-stat & P-val\\
\midrule
True & -0.177*** & 0.026 & -6.73 & 0.000\\
Qwen & 0.704* & 0.366 & 1.93 & 0.066 \\
Inattentiveness & 0.181 & 0.259 & 0.70 & 0.491 \\
Qwen*Inattentiveness & -0.787* & 0.397 & -1.98 & 0.059\\
\midrule
Intercept & -0.869*** & 0.214 & -4.07 & 0.000\\
\bottomrule
\end{tabular}
\end{table}

\section{Further Discussions}

\subsection{Robustness Test}
To ensure the robustness of our findings, we conducted several additional analyses. First, we replaced the dependent variable with Mean Absolute Error (MAE) whether our results are sensitive to different error metrics. Second, we constructed alternative interaction terms by using a continuous measure rather than the binary indicator used in our main analysis. 
\subsubsection{Alternative Dependent Variable}
We replaced the dependent variable $Y_{ij}$ with the logarithm of Mean Absolute Error (MAE) in our regression analysis. The MAE is calculated as:

\begin{equation}
MAE = \frac{1}{n}\sum_{t=1}^{n}|f_t - y_t|,
\end{equation}

where $f_t$ represents the forecast value and $y_t$ represents the actual value at time $t$. The main effects regression results demonstrate that the overall model is statistically significant when using log MAE as the dependent variable. Specifically, as shown in Table \ref{tbl:MAE_1}, the LLM-ensemble exhibits a negative and significant coefficient, suggesting that it can significantly reduce forecast errors compared to sample averaging.

Examining the regression results for each macroeconomic indicator, Table \ref{tbl:h3_gdp} reports that for GDP growth, the LLM-ensemble yields a significant improvement in one-year-ahead predictions. However, for two-year-ahead forecasts, the effect of the LLM-ensemble is smaller and not statistically significant. Similarly, Table \ref{tbl:h3_unemp} provides evidence that the LLM-ensemble offers advantages for the unemployment rate, particularly at shorter horizons.
For HICP inflation forecasts (Table \ref{tbl:h3_inf}), the coefficients for the LLM-ensemble are small and not significant.

\begin{table}[htbp]
    \caption{Comparative Forecast Accuracy Analysis with MAE (h=1)}
    \label{tbl:MAE_1}
    \begin{tabular}{@{} lccc @{}}
    \toprule
    & \multicolumn{3}{c}{APE} \\
    \cmidrule{2-4}
    & Coef. & SE & P-value \\
    \midrule
    Intercept &  -0.609** & 0.252 & 0.023 \\
    Qwen & -0.118** & 0.046 & 0.017 \\
    True & -0.609 & 0.055 & 0.247 \\
    \bottomrule
    \end{tabular}
    \end{table}

    \begin{table}[htbp]
        \caption{Comparative Forecast Accuracy Analysis of the Growth GDP with MAE}
        \label{tbl:h3_gdp}
        \begin{tabular}{@{} lcccccccc @{}}
        \toprule
        & \multicolumn{4}{c}{h=1} & \multicolumn{4}{c}{h=2} \\
        \cmidrule(lr){2-5} \cmidrule(lr){6-9}
        & Coef. & SE & t-stat & P-val & Coef. & SE & t-stat & P-val \\
        \midrule
        True & -0.142*** & 0.031 & -4.58 & 0.000 & -0.090 & 0.098 & -0.91 & 0.370 \\
        Qwen & -0.133* & 0.075 & -1.77 & 0.088 & 0.138 & 0.133 & 1.04 & 0.310 \\
        \midrule
        Intercept & -0.225 & 0.289 & -0.78 & 0.443 & -0.382* & 0.214 & -1.78 & 0.088 \\
        \bottomrule
        \end{tabular}
        \end{table}
        
        \begin{table}[htbp]
        \caption{Comparative Forecast Accuracy Analysis of the HICP Inflation with MAE}
        \label{tbl:h3_inf}
        \begin{tabular}{@{} lcccccccc @{}}
        \toprule
        & \multicolumn{4}{c}{h=1} & \multicolumn{4}{c}{h=2} \\
        \cmidrule(lr){2-5} \cmidrule(lr){6-9}
        & Coef. & SE & t-stat & P-val & Coef. & SE & t-stat & P-val \\
        \midrule
        True & 0.260** & 0.097 & 2.68 & 0.013 & 0.229** & 0.107 & 2.13 & 0.043 \\
        Qwen & -0.032 & 0.071 & -0.45 & 0.655 & -0.036 & 0.082 & -0.45 & 0.657 \\
        \midrule
        Intercept & -1.139*** & 0.309 & -3.68 & 0.001 & -1.085 & 0.322 & -3.36 & 0.003 \\
        \bottomrule
        \end{tabular}
        \end{table}
        
        \begin{table}[htbp]
        \caption{Comparative Forecast Accuracy Analysis of the Unemployment Rate with MAE}
        \label{tbl:h3_unemp}
        \begin{tabular}{@{} lcccccccc @{}}
        \toprule
        & \multicolumn{4}{c}{h=1} & \multicolumn{4}{c}{h=2} \\
        \cmidrule(lr){2-5} \cmidrule(lr){6-9}
        & Coef. & SE & t-stat & P-val & Coef. & SE & t-stat & P-val \\
        \midrule
        True & -0.466*** & 0.225 & 4.03 & 0.001 & 0.303** & 0.127 & 2.38 & 0.027 \\
        Qwen & -0.061** & 0.023 & -2.64 & 0.015 & 0.272 & 0.186 & -1.46 & 0.158\\
        \midrule
        Intercept & -3.660*** & 0.981 & -3.73 & 0.001 & -2.391** & 1.048 & -2.28 & 0.033 \\
        \bottomrule
        \end{tabular}
        \end{table}

\subsubsection{Alternative Interaction Terms}
In constructing the alternative interaction terms, we first standardize the continuous variables using $z$-score normalization, as follows:
\[
z_i = \frac{x_i - \mu_x}{\sigma_x}
\]
where $x_i$ denotes the original value, $\mu_x$ is the sample mean, and $\sigma_x$ is the sample standard deviation of the variable. The standardized variable is then multiplied by the binary indicator for Qwen to form the interaction term.

For the growth GDP, as shown in Table \ref{tbl:alt_gdp_ape} and Table \ref{tbl:alt_gdp_mae}, the interaction term exhibits a positive and statistically significant coefficient in both log(APE) and log(MAE) regressions, with estimated values of 0.513 (p=0.030) and 0.577 (p=0.026) respectively. This  indicates that the LLM-ensemble's performance deteriorates relative to the simple average when experts exhibit greater divergence in their GDP growth predictions. A plausible explanation for this finding is that expert forecasts for GDP growth inherently demonstrate substantial accuracy. Even in scenarios of high disagreement, the simple average remains effective at capturing the central tendency of these high-quality predictions.

In contrast, the unemployment rate exhibits a markedly different pattern, as evidenced in Table \ref{tbl:alt_unemp_ape}: the interaction term is negative (-0.109) and statistically significant (p=0.057), suggesting that the LLM-ensemble demonstrates superior performance when experts disagree more profoundly about unemployment dynamics. This contrasting result may be attributed to the inherently greater complexity in forecasting unemployment, where expert predictions are more susceptible to errors. In scenarios of high disagreement, the simple average becomes less reliable as a predictive tool for unemployment. Under these conditions, the LLM-ensemble leverages its sophisticated language modeling capabilities to more effectively synthesize and reconcile heterogeneous information compared to a simple arithmetic mean.

\begin{table}[htbp]
    \caption{Alternative Disagreement Interaction Analysis of the GDP Growth with log(APE) (h=2)}
    \label{tbl:alt_gdp_ape}
    \begin{tabular}{@{} lcccc @{}}
    \toprule
    & Coef. & SE & t-stat & P-val\\
    \midrule
    Qwen & 0.132 & 0.116 & 1.14 & 0.267\\
    Qwen × Disagreement & 0.513* & 0.222 & 2.30 & 0.030 \\
    \midrule
    Intercept & -0.803*** & 0.197 & -4.07 & 0.000\\
    \bottomrule
    \end{tabular}
    \end{table}
    
    \begin{table}[htbp]
    \caption{Alternative Disagreement Interaction Analysis of the GDP Growth with log(MAE) (h=2)}
        \label{tbl:alt_gdp_mae}
    \begin{tabular}{@{} lcccc @{}}
    \toprule
    & Coef. & SE & t-stat & P-val\\
    \midrule
    Qwen & 0.167 & 0.124 & 1.34 & 0.193\\
    Qwen × Disagreement & 0.577* & 0.243 & 2.37 & 0.026 \\
    \midrule
    Intercept & -0.360* & 0.203 & -1.77 & 0.090\\
    \bottomrule
    \end{tabular}
    \end{table}

\begin{table}[htbp]
\caption{Alternative Disagreement Interaction Analysis of the Unemployment Rate with log(APE) (h=1)}
\label{tbl:alt_unemp_ape}
\begin{tabular}{@{} lcccc @{}}
\toprule
& Coef. & SE & t-stat & P-val\\
\midrule
Qwen & -0.068*** & 0.022 & -3.05 & 0.006\\
Qwen × Disagreement & -0.109** & 0.054 & -2.01 & 0.057 \\
\midrule
Intercept & -4.610*** & 0.963 & -4.78 & 0.000 \\
\bottomrule
\end{tabular}
\end{table}

\begin{table}[htbp]
\caption{Alternative Disagreement Interaction Analysis of the Unemployment Rate with log(MAE) (h=1)}
\label{tbl:alt_unemp_mae}
\begin{tabular}{@{} lcccc @{}}
\toprule
& Coef. & SE & t-stat & P-val\\
\midrule
Qwen & -0.069*** & 0.023 & -3.10 & 0.003\\
Qwen × Disagreement & -0.109** & 0.054 & -2.01 & 0.057 \\
\midrule
Intercept & -3.55*** & 0.965 & -3.68 & 0.001\\
\bottomrule
\end{tabular}
\end{table}

\subsubsection{Alternative LLM Model}

To test the robustness of our findings, we replace Qwen with Deepseek as an alternative LLM model \citep{bi2024deepseek}. Tables \ref{tbl:ape_deepseek} and \ref{tbl:mae_deepseek} present the main effects analysis using both logrithm of APE and MAE as dependent variables. The results show that Deepseek-ensemble achieves significantly better forecast accuracy compared to simple averaging for one-quarter-ahead forecasts.

We further examine specific scenarios to validate the robustness of our findings. For the unemployment rate forecasts, we test both binary and continuous forms of expert disagreement interaction terms, as shown in Table \ref{tbl:disagreement_ape}. Under the binary interaction specification, Deepseek-ensemble effectively reduces forecast errors compared to simple averaging when disagreement was higher. When using continuous disagreement measures, the interaction term remained negative (-0.046) but became statistically insignificant (p = 0.196). The consistent negative direction of both interaction specifications suggests that the LLM-ensemble's ability to reconcile divergent expert opinions is generally robust, though the strength of this effect may vary depending on how disagreement is measured. The LLM-ensemble demonstrates robust consistency in reconciling divergent expert opinions; however, the magnitude of this reconciliation effect exhibits sensitivity to the choice of disagreement metric, suggesting model-agnostic stability in consensus formation.

Similarly, we test the robustness of expert inattentiveness effects using both binary and continuous specifications, as shown in Table \ref{tbl:attention_ape}. Under both situations, Deepseek-ensemble shows improved performance with a negative interaction coefficient. This pattern indicates that while LLM-ensemble generally maintains an advantage in managing expert inattentiveness, the strength of this effect depends on measurement approach. The weaker significance in continuous specification may stem from the gradual nature of attention changes being harder to capture than discrete shifts. These findings reinforce that LLM-ensemble's ability to compensate for expert inattentiveness is fundamentally robust.The detailed regression results are presented in the Appendix.

\begin{table}[htbp]
    \caption{Comparative Forecast Accuracy Analysis with APE by Deepseek}
    \label{tbl:ape_deepseek}
    \begin{tabular}{@{}lccc|ccc@{}}
    \toprule
    & \multicolumn{3}{c|}{h=1} & \multicolumn{3}{c}{h=2} \\
    \cmidrule{2-4} \cmidrule{5-7}
    & Coef. & SE & P-value & Coef. & SE & P-value \\
    \midrule
    Deepseek & -0.081* & 0.046 & 0.092 & 0.126 & 0.077 & 0.117 \\
    True & -0.149*** & 0.027 & 0.000 & -0.185*** & 0.027 & 0.000 \\
    Intercept & -0.788*** & 0.152 & 0.000 & -0.666*** & 0.157 & 0.000 \\
    \bottomrule
    \end{tabular}
    \end{table}
    
    \begin{table}[htbp]
    \caption{Comparative Forecast Accuracy Analysis with MAE by Deepseek}
    \label{tbl:mae_deepseek}
    \begin{tabular}{@{}lccc|ccc@{}}
    \toprule
    & \multicolumn{3}{c|}{h=1} & \multicolumn{3}{c}{h=2} \\
    \cmidrule{2-4} \cmidrule{5-7}
    & Coef. & SE & P-value & Coef. & SE & P-value \\
    \midrule
    Deepseek & -0.079* & 0.046 & 0.098 & 0.137* & 0.082 & 0.108 \\
    True & 0.072 & 0.057 & 0.214 & 0.036 & 0.055 & 0.521 \\
    Intercept & -0.631** & 0.257 & 0.021 & -0.513* & 0.248 & 0.050 \\
    \bottomrule
    \end{tabular}
    \end{table}

    \begin{table}[htbp]
        \caption{Different Interaction Forms of the Disagreement of the Unemployment Rate with log(APE) (h=1)}
        \label{tbl:disagreement_ape}
        \begin{tabular}{@{}lccc|ccc@{}}
        \toprule
        & \multicolumn{3}{c|}{Binary} & \multicolumn{3}{c}{Continuous} \\
        \cmidrule{2-4} \cmidrule{5-7}
        & Coef. & SE & P-value & Coef. & SE & P-value \\
        \midrule
        True & 0.286** & 0.106 & 0.013 & 0.290*** & 0.096 & 0.006 \\
        Deepseek & 0.182* & 0.080 & 0.032 & 0.074 & 0.046 & 0.126 \\
        Interaction & -0.240* & 0.102 & 0.028 & -0.046 & 0.034 & 0.196 \\
        Intercept & -4.468*** & 0.920 & 0.000 & -4.361*** & 0.821 & 0.000 \\
        \bottomrule
        \end{tabular}
        \end{table}

        \begin{table}[htbp]
            \caption{Different Interaction Forms of the Inattentiveness of the HICP Inflation with log(APE) (h=1)}
            \label{tbl:attention_ape}
            \begin{tabular}{@{}lccc|ccc@{}}
            \toprule
            & \multicolumn{3}{c|}{Binary} & \multicolumn{3}{c}{Continuous} \\
            \cmidrule{2-4} \cmidrule{5-7}
            & Coef. & SE & P-value & Coef. & SE & P-value \\
            \midrule
            True & -0.163 & 0.203 & 0.428 & -0.172 & 0.198 & 0.391 \\
            Deepseek & -0.180** & 0.064 & 0.009 & 0.123 & 0.097 & 0.214 \\
            Interaction & -1.308** & 0.497 & 0.014 & -0.418* & 0.182 & 0.030 \\
            Intercept & -0.580 & 0.548 & 0.300 & -0.655 & 0.519 & 0.218 \\
            \bottomrule
            \end{tabular}
            \end{table}

\subsection{Ablation Study}
To investigate the impact of different prompt components on the LLM-ensemble performance, we conduct an ablation study by systematically removing each dimension from the prompt design. Specifically, we examine three key dimensions from our prompt engineering framework: (1) historical accuracy weighting, (2) lag compensation, and (3) trend enhancement. For each reduced prompt, we run the main effects regression for one-year-ahead forecasts with the logarithm of APE and MAE as dependent variables. Tables \ref{tbl:p1}, \ref{tbl:p2}, and \ref{tbl:p3} present the regression results when removing each component respectively. The results show that removing historical accuracy weighting leads to a smaller coefficient (-0.041) with increased standard error (0.035). Similarly, excluding lag compensation results in a diminished coefficient (-0.030), while removing trend enhancement yields the weakest performance (-0.028). The consistently negative but insignificant coefficients across all three reduced prompts, compared to the significant improvements shown in our main results, demonstrate that all three dimensions are essential for optimal LLM-ensemble performance. This finding validates our comprehensive prompt engineering approach that integrates historical performance analysis, temporal calibration, and trend incorporation.

\begin{table}[htbp]
    \caption{Comparative Forecast Accuracy Analysis (h=1) - Prompt 1}
    \label{tbl:p1}
    \begin{tabular}{@{}lccc|ccc@{}}
    \toprule
    & \multicolumn{3}{c|}{log(APE)} & \multicolumn{3}{c}{log(MAE)} \\
    \cmidrule{2-4} \cmidrule{5-7}
    & Coef. & SE & P-value & Coef. & SE & P-value \\
    \midrule
    Intercept & -0.759*** & 0.152 & 0.000 & -0.602** & 0.254 & 0.026 \\
    Qwen & -0.041 & 0.035 & 0.258 & -0.035 & 0.035 & 0.326 \\
    True & -0.157*** & 0.027 & 0.000 & 0.064 & 0.056 & 0.265 \\
    \bottomrule
    \end{tabular}
    \end{table}
    
    \begin{table}[htbp]
    \caption{Comparative Forecast Accuracy Analysis (h=1) - Prompt 2}
    \label{tbl:p2}
    \begin{tabular}{@{}lccc|ccc@{}}
    \toprule
    & \multicolumn{3}{c|}{log(APE)} & \multicolumn{3}{c}{log(MAE)} \\
    \cmidrule{2-4} \cmidrule{5-7}
    & Coef. & SE & P-value & Coef. & SE & P-value \\
    \midrule
    Intercept & -0.759*** & 0.150 & 0.000 & -0.606** & 0.248 & 0.022 \\
    Qwen & -0.030 & 0.030 & 0.335 & -0.031 & 0.031 & 0.315 \\
    True & -0.157*** & 0.026 & 0.000 & 0.065 & 0.054 & 0.242 \\
    \bottomrule
    \end{tabular}
   
    \end{table}
    
    \begin{table}[htbp]
    \caption{Comparative Forecast Accuracy Analysis (h=1) - Prompt 3}
    \label{tbl:p3}
    \begin{tabular}{@{}lccc|ccc@{}}
    \toprule
    & \multicolumn{3}{c|}{log(APE)} & \multicolumn{3}{c}{log(MAE)} \\
    \cmidrule{2-4} \cmidrule{5-7}
    & Coef. & SE & P-value & Coef. & SE & P-value \\
    \midrule
    Intercept & -0.760*** & 0.150 & 0.000 & -0.607** & 0.248 & 0.022 \\
    Qwen & -0.028 & 0.030 & 0.364 & -0.030 & 0.031 & 0.340 \\
    True & -0.157*** & 0.026 & 0.000 & 0.065 & 0.054 & 0.239 \\
    \bottomrule
    \end{tabular}
    \end{table}

\subsection{Interpretation of LLM-Ensemble}

To better understand the reasoning process behind the forecast combination generated by LLMs, we prompted the LLM to explicitly articulate its logic for assigning expert weights and synthesizing predictions. As an illustrative example, we present below the LLM's step-by-step reasoning for the ensemble forecast in 1999Q4.

\begin{verbatim}

In generating the ensemble forecast for 1999Q4, we first evaluated each expert's 
historical prediction accuracy by examining the deviation between their forecasts 
for 1999Q3 and the actual value of 3.1. 
Experts 4 and 53 exhibited substantial prediction errors in 1999Q3, 
and thus their weights were correspondingly reduced in the 1999Q4 forecast. 
Simultaneously, considering the presence of lag effects in expert predictions, 
we adjusted the weights of those experts who demonstrated lagged responses in 1999Q3 
(such as Experts 71 and 72). 
Furthermore, by analyzing the trend of the actual value in 1999Q3, 
which was relatively stable, 
we assigned higher weights to experts whose predictions were closer to the true value. 
Experts 4 and 53 received lower weights, Experts 71 and 72 were also adjusted, 
while those whose predictions were close to the actual value in 1999Q3 were assigned 
higher weights.
Ultimately, by integrating these factors, the ensemble forecast for 1999Q4 was 
determined to be 2.4.
\end{verbatim}

The above reasoning reveals that the LLM-ensemble employs a multi-stage, cognitively informed approach to expert weighting and forecast synthesis:

\begin{enumerate}
    \item \textbf{Error-Driven Weighting:} The model initially prioritizes experts based on their recent forecasting accuracy, systematically reducing the influence of those with larger errors (e.g., Expert 4 and 53). This dynamic adjustment is consistent with Bayesian updating principles and established practices in forecast combinations.
    \item \textbf{Lag Effect Identification and Correction:} The ensemble explicitly detects lagged behavioral patterns among experts (e.g., Expert 71 and 72) and proactively adjusts their weights to mitigate the impact of systematic delays. This mechanism parallels human strategies for correcting temporal biases in time series analysis.
    \item \textbf{Trend Consistency Optimization:} By referencing the stability of the actual value in 1999Q3, the model further refines its weighting scheme to favor experts whose predictions align closely with observed trends. This reflects a logic that balances persistence with adaptabilit in macroeconomic forecasting.
\end{enumerate}

Synthesizing these elements, the LLM-ensemble demonstrates a hierarchical reasoning process: it begins with quantitative error analysis, incorporates qualitative behavioral adjustments, and finally assigns weights according to macro-level trends. The combiantion of both statistical rigor and behavioral insight positions the LLM-ensemble as an advancement, bridging the gap between algorithmic combination and expert judgment in economic forecasting.

\subsection{Emotional Impacts of LLM}

To explore how emotional prompt affects LLM-ensemble, we incorporated sentiment orientations (optimistic, neutral, and pessimistic) into the prompting process. Given that macroeconomic forecasting is inherently tied to market sentiment and expert expectations, we hypothesized that emotional context could help capture different aspects of forecaster behavior and market dynamics. By prefacing the forecast combination task with sentiment-specific framing "You have a more positive/neutral/negative outlook on the future market.", we examined how different emotional contexts influence the ensemble weights and resulting predictions.

\begin{figure}[htbp]
\centering
\includegraphics[width=\linewidth]{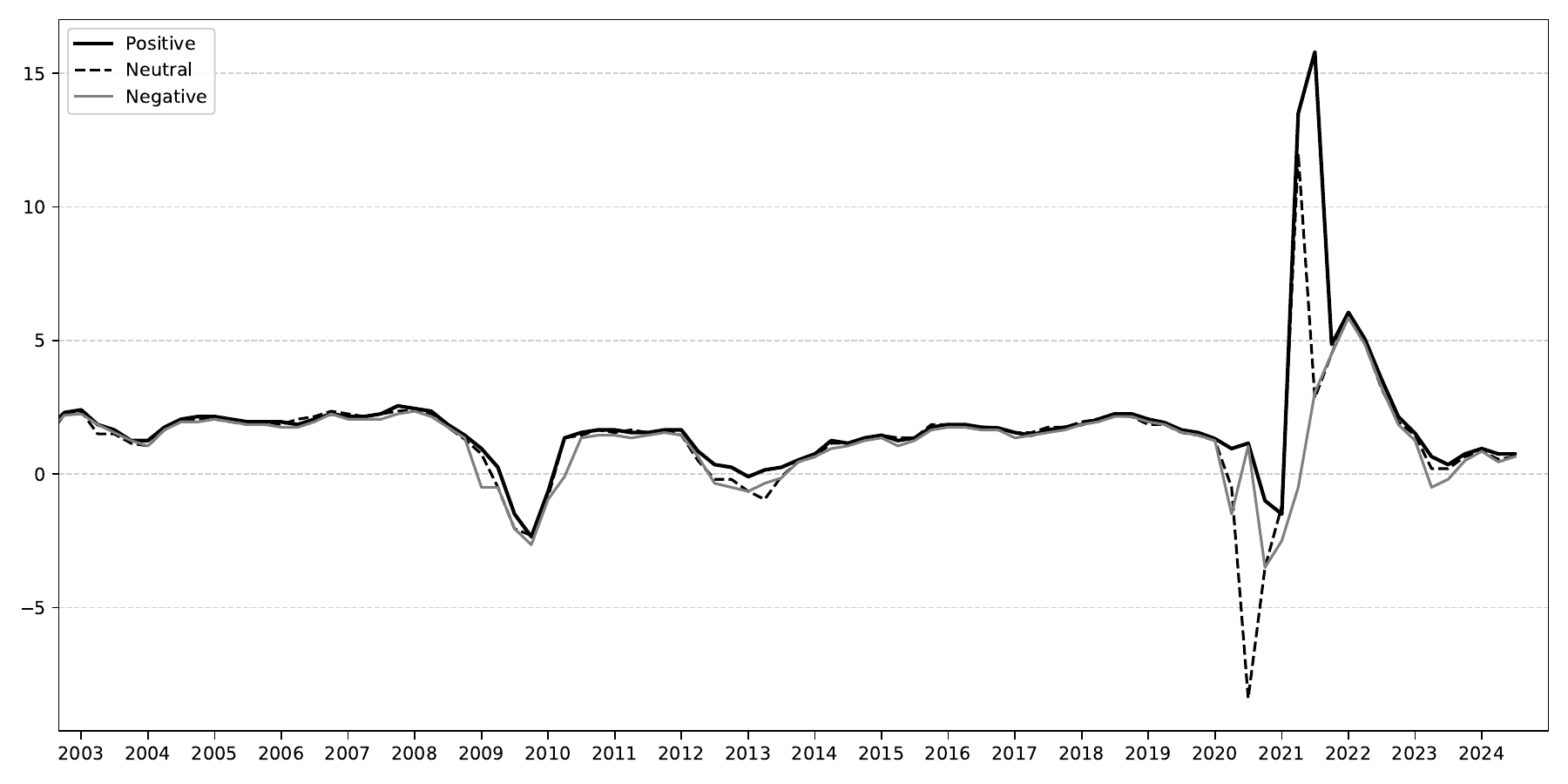}
\caption{GDP Growth Emsemble Results under Different Sentiment Orientations}
\label{fig:gdp_sent}
\end{figure}

\begin{figure}[htbp]
\centering
\includegraphics[width=\linewidth]{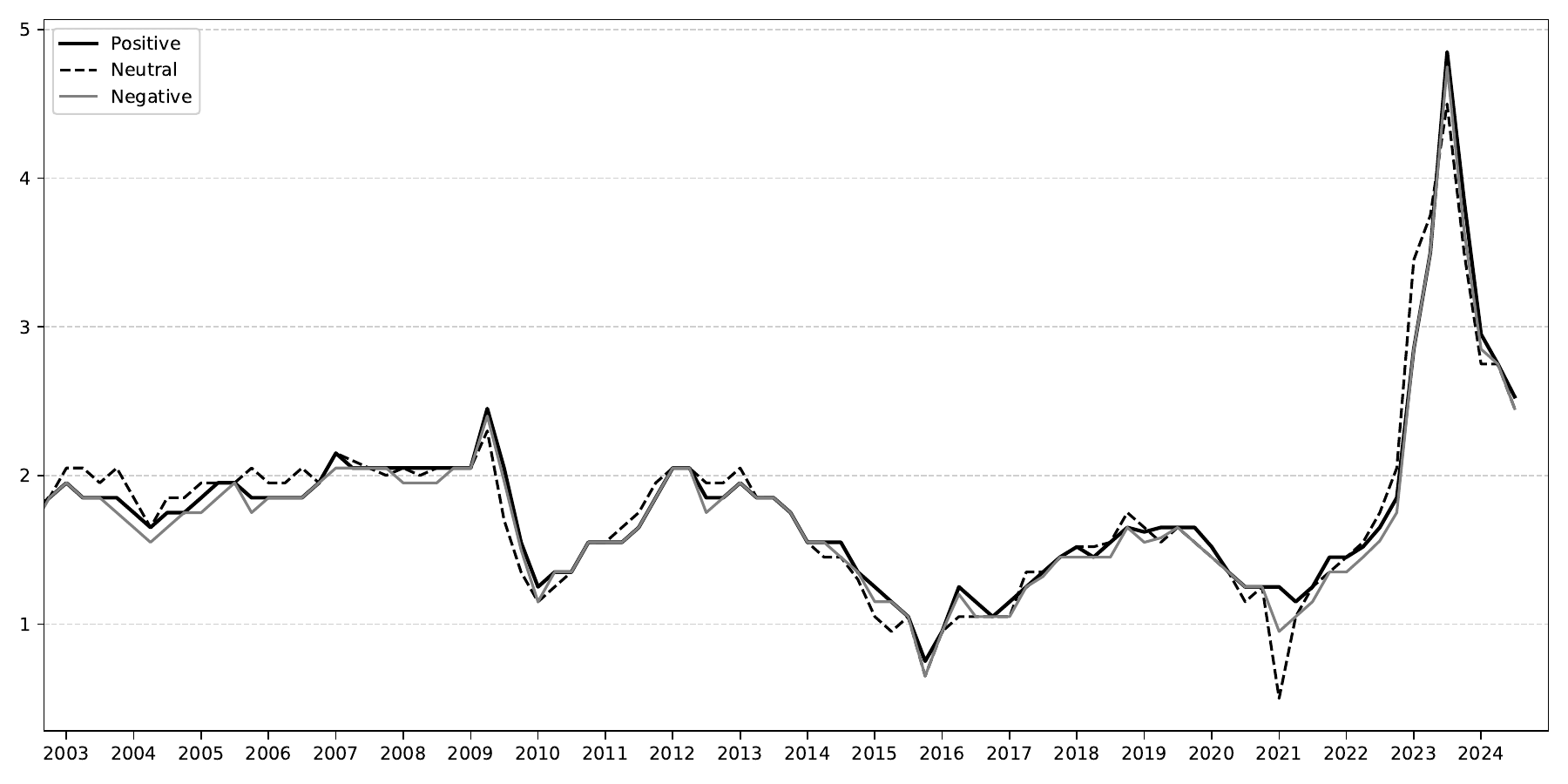}
\caption{HICP Inflation Emsemble Results under Different Sentiment Orientations}            
\label{fig:inf_sent}
\end{figure}

\begin{figure}[htbp]
\centering
\includegraphics[width=\linewidth]{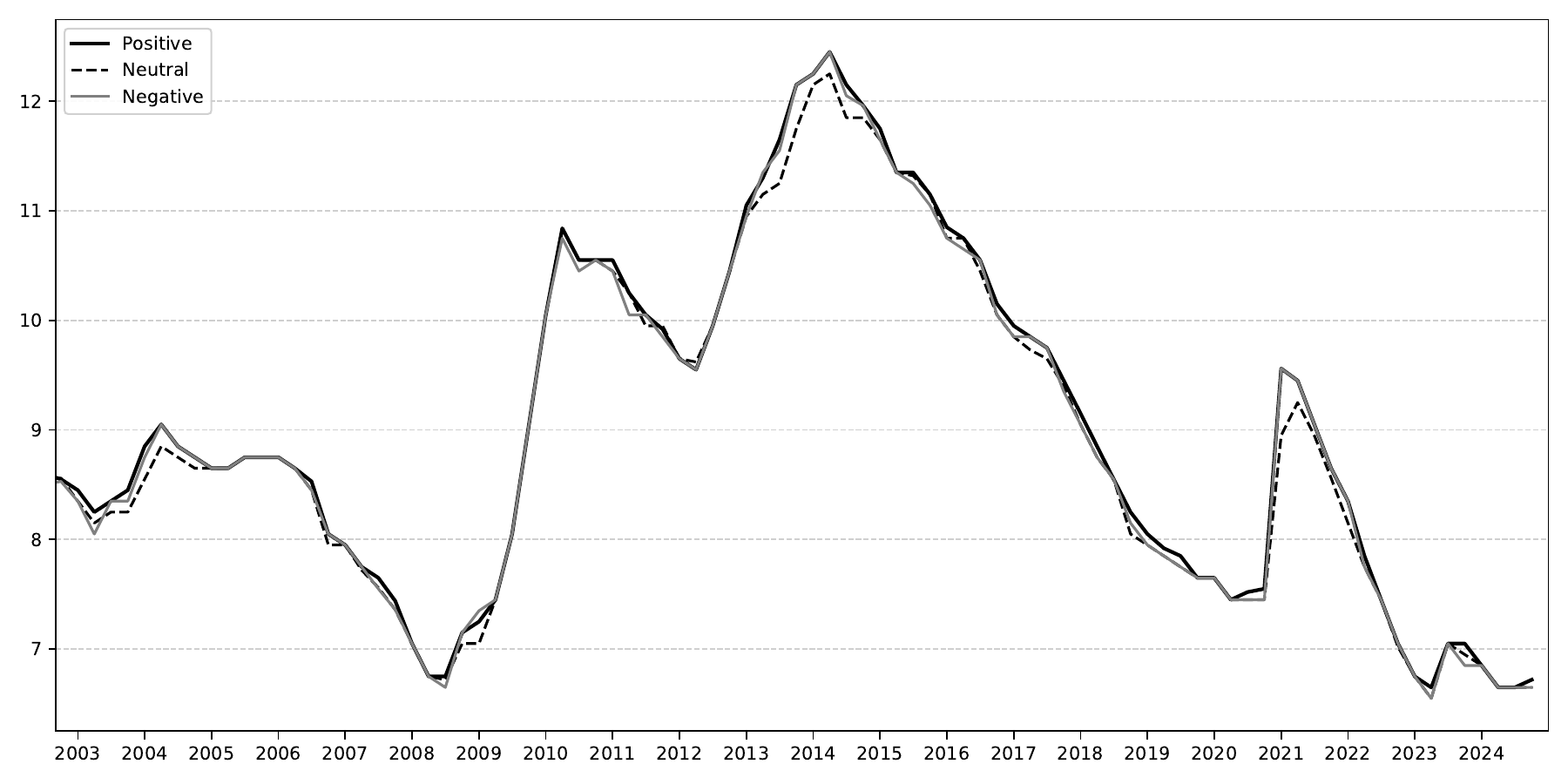}
\caption{Unemployment Rate Emsemble Results under Different Sentiment Orientations}
\label{fig:unemp_sent}
\end{figure}

As shown in figures above, the impact of emotional prompt varies systematically across different macroeconomic variables. For GDP growth, optimistic scenarios generate notably higher predictions during economic turning points, particularly in periods of sharp transitions, and has enhanced sensitivity to positive momentum signals. In contrast, inflation forecasts exhibit an inverse relationship, with pessimistic scenarios producing higher predictions - a pattern that aligns with how inflation concerns typically correlate with negative market sentiment. The unemployment rate demonstrates the most intuitive sentiment-driven pattern, with optimistic scenarios consistently yielding lower predictions than neutral or pessimistic cases.

The systematic variations across indicators demonstrate that sentiment-aware ensemble methods can enhance differences in expert predictions. More importantly, by incorporating different emotional prompts into LLM inputs, we can generate more diverse and comprehensive forecasting results that better reflect the full spectrum of market sentiment and expert expectations.

\section{Conclusion}

This paper investigates the potential of leveraging large language models (LLMs) for macroeconomic forecast combination, motivated by the need to overcome limitations in traditional averaging methods and harness LLMs' capabilities in processing complex information. Through systematic empirical analysis of ECB Survey of Professional Forecasters data, we find that the LLM-ensemble approach demonstrates superior performance compared to simple averaging across multiple indicators, particularly for one-year-ahead forecasts. Specifically, our results reveal that LLM-ensemble exhibits greater resilience to expert disagreement (H3) and maintains relatively stable performance under varying levels of forecaster inattentiveness (H4).

The key innovation of our approach lies in the development of a zero-shot learning framework that requires no pre-training or parameter tuning. Through carefully engineered prompts incorporating historical accuracy weighting, lag compensation, and trend enhancement mechanisms, we enable LLMs to dynamically analyze expert forecasting patterns and optimally combine predictions. The LLM's reasoning process demonstrates sophisticated cognitive capabilities in detecting behavioral biases, temporal lags, and systematic errors in expert forecasts.

Our findings have important implications for both research and practice. For researchers, we provide a novel methodological framework that bridges statistical and behavioral approaches to forecast combination. For practitioners, we offer an immediately deployable solution that can enhance forecast accuracy without requiring extensive model training or parameter optimization.

Future research could extend this work in several directions: (1) exploring the application of our framework to high-frequency financial forecasting, (2) investigating the potential for incorporating additional contextual information such as news sentiment and policy changes, and (3) developing more sophisticated prompt engineering techniques to further improve forecast combination accuracy. Additionally, examining how different LLM architectures affect combination performance could yield valuable insights for model selection in practical applications.

\section{Acknowledgements}

\appendix
\section{Appendix}

\begin{table}[htbp]
    \caption{Comparative Forecast Accuracy Analysis of Three Macro Indicators with Deepseek(h=1)}
    \label{tbl:log_ape_macro}
    \begin{tabular}{@{} lccc @{}}
    \toprule
    & \multicolumn{3}{c}{log(APE)} \\
    \cmidrule{2-4}
    & Coef. & SE & P-value \\
    \midrule
    \multicolumn{4}{l}{Growth GDP} \\
    Intercept & -0.779*** & 0.196 & 0.001 \\
    Deepseek & -0.264** & 0.115 & 0.030 \\
    True & -0.156*** & 0.041 & 0.001 \\
    \midrule
    \multicolumn{4}{l}{HICP Inflation} \\
    Intercept & -0.612 & 0.546 & 0.273 \\
    Deepseek & -0.027 & 0.047 & 0.566 \\
    True & -0.176 & 0.203 & 0.396 \\
    \midrule
    \multicolumn{4}{l}{Unemployment Rate} \\
    Intercept & -4.431*** & 0.898 & 0.000 \\
    Deepseek & 0.059 & 0.040 & 0.162 \\
    True & 0.302*** & 0.105 & 0.009 \\
    \bottomrule
    \end{tabular}
    \end{table}

\begin{table}[htbp]
    \caption{Impact of Expert Disagreement on Forecast Methods with Deepseek of log(APE)}
    \label{tbl:log_ape_disagreement}
    \begin{tabular}{@{} lccccccc @{}}
    \toprule
    & \multicolumn{3}{c}{h=1} & \multicolumn{3}{c}{h=2} \\
    \cmidrule{2-4} \cmidrule{5-7}
    & Coef. & SE & P-value & Coef. & SE & P-value \\
    \midrule
    Intercept & -0.938*** & 0.189 & 0.000 & -0.818*** & 0.206 & 0.001 \\
    Deepseek & 0.026 & 0.123 & 0.836 & 0.144 & 0.133 & 0.289 \\
    Deepseek*Disagreement & -0.203 & 0.220 & 0.366 & -0.021 & 0.192 & 0.914 \\
    \bottomrule
    \end{tabular}
    \end{table}

\begin{table}[htbp]
    \caption{Impact of Expert Disagreement on Forecast Methods with Deepseek of log(MAE)}
    \label{tbl:log_mae_disagreement}
    \begin{tabular}{@{} lccccccc @{}}
    \toprule
    & \multicolumn{3}{c}{h=1} & \multicolumn{3}{c}{h=2} \\
    \cmidrule{2-4} \cmidrule{5-7}
    & Coef. & SE & P-value & Coef. & SE & P-value \\
    \midrule
    Intercept & -0.714** & 0.292 & 0.022 & -0.637* & 0.293 & 0.040 \\
    Deepseek & -0.013 & 0.107 & 0.907 & 0.152 & 0.126 & 0.239 \\
    Deepseek*Disagreement & -0.127 & 0.183 & 0.494 & -0.018 & 0.184 & 0.923 \\
    \bottomrule
    \end{tabular}
    \end{table}

\begin{table}[htbp]
    \caption{Impact of Expert Inattentiveness on Forecast Methods with Deepseek of log(APE)}
    \label{tbl:log_ape_inattentive}
    \begin{tabular}{@{} lccccccc @{}}
    \toprule
    & \multicolumn{3}{c}{h=1} & \multicolumn{3}{c}{h=2} \\
    \cmidrule{2-4} \cmidrule{5-7}
    & Coef. & SE & P-value & Coef. & SE & P-value \\
    \midrule
    Intercept & -0.795*** & 0.167 & 0.000 & -0.771*** & 0.175 & 0.000 \\
    Deepseek & -0.147* & 0.083 & 0.091 & 0.084 & 0.095 & 0.381 \\
    Deepseek*Inattentiveness & 0.235 & 0.203 & 0.256 & 0.125 & 0.220 & 0.576 \\
    \bottomrule
    \end{tabular}
    \end{table}

\begin{table}[htbp]
    \caption{Impact of Expert Inattentiveness on Forecast Methods with Deepseek of log(MAE)}
    \label{tbl:log_mae_inattentive}
    \begin{tabular}{@{} lccccccc @{}}
    \toprule
    & \multicolumn{3}{c}{h=1} & \multicolumn{3}{c}{h=2} \\
    \cmidrule{2-4} \cmidrule{5-7}
    & Coef. & SE & P-value & Coef. & SE & P-value \\
    \midrule
    Intercept & -0.658** & 0.242 & 0.012 & -0.658*** & 0.233 & 0.009 \\
    Deepseek & -0.110 & 0.086 & 0.210 & 0.150 & 0.095 & 0.128 \\
    Deepseek*Inattentiveness & 0.105 & 0.216 & 0.629 & -0.073 & 0.200 & 0.720 \\
    \bottomrule
    \end{tabular}
    \end{table}

\begin{table}[htbp]
    \caption{Impact of Contimous Intersec of Disagreement on Forecast Methods with Deepseek of log(APE)}
    \label{tbl:log_ape_continuous_disagreement}
    \begin{tabular}{@{} lccccccc @{}}
    \toprule
    & \multicolumn{3}{c}{1-year} & \multicolumn{3}{c}{2-year} \\
    \cmidrule{2-4} \cmidrule{5-7}
    & Coef. & SE & P-value & Coef. & SE & P-value \\
    \midrule
    Intercept & -0.782*** & 0.149 & 0.000 & -0.649*** & 0.156 & 0.000 \\
    Deepseek & -0.077 & 0.047 & 0.112 & 0.140 & 0.075 & 0.075 \\
    Intersection & -0.002 & 0.067 & 0.977 & 0.109 & 0.099 & 0.283 \\
    \bottomrule
    \end{tabular}
    \end{table}

\begin{table}[htbp]
    \caption{Impact of Contimous Intersec of Disagreement on Forecast Methods with Deepseek of log(MAE)}
    \label{tbl:log_mae_continuous_disagreement}
    \begin{tabular}{@{} lccccccc @{}}
    \toprule
    & \multicolumn{3}{c}{h=1} & \multicolumn{3}{c}{h=2} \\
    \cmidrule{2-4} \cmidrule{5-7}
    & Coef. & SE & P-value & Coef. & SE & P-value \\
    \midrule
    Intercept & -0.623** & 0.255 & 0.022 & -0.495* & 0.245 & 0.055 \\
    Deepseek & -0.074 & 0.047 & 0.124 & 0.153 & 0.080 & 0.068 \\
    Intersection & 0.047 & 0.047 & 0.326 & 0.105 & 0.090 & 0.254 \\
    \bottomrule
    \end{tabular}
    \end{table}

\begin{table}[htbp]
    \caption{Impact of Contimous Intersec of Inattentiveness on Forecast Methods with Deepseek of log(APE)}
    \label{tbl:log_ape_continuous_inattentive}
    \begin{tabular}{@{} lccccccc @{}}
    \toprule
    & \multicolumn{3}{c}{h=1} & \multicolumn{3}{c}{h=2} \\
    \cmidrule{2-4} \cmidrule{5-7}
    & Coef. & SE & P-value & Coef. & SE & P-value \\
    \midrule
    Intercept & -1.125 & 0.783 & 0.163 & -1.368 & 0.816 & 0.107 \\
    Deepseek & -0.949 & 0.786 & 0.239 & -0.620 & 0.858 & 0.477 \\
    Intersection & 1.050 & 0.943 & 0.276 & 0.927 & 1.040 & 0.381 \\
    \bottomrule
    \end{tabular}
    \end{table}

\begin{table}[htbp]
    \caption{Impact of Contimous Intersec of Inattentiveness on Forecast Methods with Deepseek of log(MAE)}
    \label{tbl:log_mae_continuous_inattentive}
    \begin{tabular}{@{} lccccccc  @{}}
    \toprule
    & \multicolumn{3}{c}{h=1} & \multicolumn{3}{c}{h=2} \\
    \cmidrule{2-4} \cmidrule{5-7}
    & Coef. & SE & P-value & Coef. & SE & P-value \\
    \midrule
    Intercept & -1.451** & 0.604 & 0.024 & -1.913*** & 0.602 & 0.004 \\
    Deepseek & -0.156 & 0.691 & 0.823 & 0.102 & 0.739 & 0.892 \\
    Intersection & 0.082 & 0.822 & 0.921 & 0.060 & 0.898 & 0.948 \\
    \bottomrule
    \end{tabular}
    \end{table}

\bibliographystyle{cas-model2-names}

\bibliography{cas-refs}


\end{document}